\documentclass[onecolumn,superscriptaddress,notitlepage, nofootinbib]{revtex4-1}
\pdfoutput=1

\usepackage{graphicx}
\usepackage{amsmath}
\usepackage{hyperref}
\usepackage[utf8]{inputenc}
\usepackage[dvipsnames]{xcolor}
\usepackage{subfigure}

\usepackage{dcolumn}

\hypersetup{colorlinks,linkcolor={blue},citecolor={teal},urlcolor={violet}}

\newcommand{\GRAPPA}{GRAPPA Institute, University of Amsterdam, 1098 XH Amsterdam, The Netherlands}
\newcommand{\KAVLI}{Kavli Institute for the Physics and Mathematics of the Universe (Kavli IPMU, WPI), University of Tokyo, Kashiwa, Chiba 277-8583, Japan}

\begin{document}

\title{Probing dark matter signals in neutrino telescopes through angular power spectrum}

\author{Ariane Dekker}
\email{a.h.dekker@uva.nl}
\affiliation{\GRAPPA}

\author{Marco Chianese}
\email{m.chianese@uva.nl}
\affiliation{\GRAPPA}

\author{Shin’ichiro Ando}
\email{s.ando@uva.nl}
\affiliation{\GRAPPA}
\affiliation{\KAVLI}

\date{\today}

\begin{abstract}
The hypothesis of two different components in the high-energy neutrino flux observed with IceCube has been proposed to solve the tension among different data-sets and to account for an excess of neutrino events at 100~TeV. In addition to a standard astrophysical power-law component, the second component might be explained by a different class of astrophysical sources, or more intriguingly, might originate from decaying or annihilating dark matter. These two scenarios can be distinguished thanks to the different expected angular distributions of neutrino events. Neutrino signals from dark matter are indeed expected to have some correlation with the extended galactic dark matter halo. In this paper, we perform angular power spectrum analyses of simulated neutrino sky maps to investigate the two-component hypothesis with a contribution from dark matter. We provide current constraints and expected sensitivity to dark matter parameters for future neutrino telescopes such as IceCube-Gen2 and KM3NeT. The latter is found to be more sensitive than IceCube-Gen2 to look for a dark matter signal at low energies towards the galactic center. Finally, we show that after 10 years of data-taking, they will firmly probe the current best-fit scenario for decaying dark matter by exploiting the angular information only.
\end{abstract}

\maketitle


\section{Introduction}

Astrophysical high-energy neutrinos were detected for the first time with IceCube in the last years~\cite{Aartsen:2013jdh,Aartsen:2014gkd,Aartsen:2017mau}. Since then, IceCube~\cite{Schneider:2019ayi,Stettner:2019tok} and ANTARES~\cite{Albert:2017nsd,ANTARESicrc19} are collecting more neutrino events with energies between TeV and PeV, confirming the presence of an extraterrestrial neutrino flux. These high-energy neutrinos are expected to be produced within or in the surroundings of cosmic-ray accelerators in hadronic~\cite{Loeb:2006tw} or photo-hadronic~\cite{Winter:2013cla} interactions through the decay of charged pions. Neutrinos travel in an unattenuated and undeviated path towards Earth and can therefore provide insight into acceleration processes, on the origin of high-energy cosmic rays and on the potential discovery of new distant sources. Moreover, along with neutrinos, gamma rays are also produced in cosmic-ray accelerators through the decay of neutral pions. Detecting both neutrinos and gamma rays from the same source would be compelling evidence for hadronic interactions, and in fact, the flaring blazar TXS 0506+056 showed temporal and spatial coincidence with neutrino emission~\cite{Ahnen:2018mvi,IceCube:2018dnn,IceCube:2018cha}. However, it was also found that blazars can only contribute up to $10\%$~\cite{Aartsen:2016lir,Murase:2018iyl}, and the resulting sources observed by IceCube remain unknown. Point-like searches correlating gamma-ray source catalogs with neutrino data~\cite{Adrian-Martinez:2015ver,Aartsen:2016oji,Aartsen:2018fpd,Aartsen:2018ywr,Carver:2019jcd}, as well as angular clustering analysis~\cite{Aartsen:2014hva,Aartsen:2014ivk,Mertsch:2016hcd,Murase:2016gly,Ando:2017xcb,Dekker:2018cqu, Aartsen:2019mbc, Aartsen:2019fau}, have both led to upper limits on the contribution of source classes and on the flux per individual source. In particular, Ref.~\cite{Dekker:2018cqu} points out that by using only isotropic and anisotropic information of the diffuse neutrino sky, upper limits can already be set on the contribution of rare and bright source classes with two years of IceCube through-going events.

If neutrinos are products of cosmic-ray accelerators, the neutrino spectrum is expected to follow a power-law, derived from the first-order Fermi shock acceleration as being proportional to $E_\nu^{-\gamma}$, with $2.0 \lesssim \gamma \lesssim 2.2$ \cite{Waxman:1998yy}. Furthermore, combining with gamma-ray data from Fermi-LAT observations, upper limits are put on the spectral index for the hadronic scenario of $\gamma\lesssim 2.1-2.2$ \cite{Murase:2013rfa,Tamborra:2014xia,Ando:2015bva,Bechtol:2015uqb}. This is in a good agreement with the recent best-fit $\gamma^{\rm TG} = 2.28$ of through-going (TG) muon neutrinos with 10 years of IceCube observation~\cite{Stettner:2019tok}. On the other hand, the 7.5-year High-Energy Starting Events (HESE) data-set prefers a steeper spectrum with a best-fit index of $\gamma^{\rm HESE}=2.89$~\cite{Schneider:2019ayi}. These two measurements are slightly in tension, suggesting that the observed neutrino spectrum does not fit very well with the assumption of a single power-law component. Remarkably, such a tension is strengthened by combining IceCube and ANTARES data~\cite{Chianese:2017jfa}. It is worth noticing that the TG data-set observes only the northern sky above 200~TeV, while the HESE data contains astrophysical neutrinos from the full sky, including the galactic center, with an energy larger than about 60~TeV. This motivates a two-component scenario, one extragalactic isotropic component with $\gamma \simeq 2.0$, and a second, possibly galactic, component with a steeper spectrum~\cite{Chen:2014gxa,Aartsen:2015knd,Chianese:2016opp,Palladino:2016zoe,Vincent:2016nut,Palladino:2016xsy,Anchordoqui:2016ewn,Palladino:2018evm,Sui:2018bbh}. The latter is required to account for an excess in neutrino events at around 100~TeV~\cite{Chianese:2016opp,Chianese:2016kpu,Chianese:2017jfa,Chianese:2017nwe}. Different interpretations of such an excess have been proposed so far: hidden astrophysical sources where gamma-rays are highly absorbed due to the source's opacity~\cite{Kimura:2014jba,Murase:2015xka,Tamborra:2015fzv,Senno:2015tsn,Denton:2017jwk,Denton:2018tdj}, invisible decay of active neutrinos~\cite{Denton:2018aml}, or decaying dark matter (DM) particles~\cite{Chianese:2016opp,Chianese:2016kpu,Chianese:2017nwe,Chianese:2018ijk,Chianese:2019kyl}.

A detectable flux of high-energy neutrinos might indeed originate from the decay of heavy DM particles~\cite{Chianese:2016opp,Chianese:2016kpu,Chianese:2017nwe,Anisimov:2008gg,Feldstein:2013kka,Esmaili:2013gha,Bai:2013nga,Ema:2013nda,Bhattacharya:2014vwa,Higaki:2014dwa,Ema:2014ufa,Rott:2014kfa,Esmaili:2014rma,Fong:2014bsa,Dudas:2014bca,Kopp:2015bfa,Murase:2015gea,Aisati:2015vma,Anchordoqui:2015lqa,Boucenna:2015tra,Ko:2015nma,Dev:2016uxj,Fiorentin:2016avj,Dev:2016qbd,DiBari:2016guw,Chianese:2016smc,Borah:2017xgm,Hiroshima:2017hmy,Bhattacharya:2017jaw,Sui:2018bbh,Chianese:2018ijk,Bhattacharya:2019ucd,Pandey:2019cre,Chianese:2019kyl,Pandey:2019wfk}. On the other hand, a neutrino signal from annihilating DM particles is in general suppressed by the unitarity limit, except in case of very cold DM substructures~\cite{Feldstein:2013kka,Zavala:2014dla,ElAisati:2017ppn,Bhattacharya:2019ucd}. Current neutrino and gamma-ray data place constraints on both decaying and annihilating DM scenarios~\cite{Albert:2016emp,Abeysekara:2017jxs,Aartsen:2018mxl,Arguelles:2019boy,Iovine:2019rmd}. Multi-messenger analyses based on the observations of neutrinos and gamma-rays are indeed of paramount importance when examining potential heavy DM signals~\cite{Cohen:2016uyg,Neronov:2018ibl,Blanco:2018esa,Ishiwata:2019aet}. In particular, the Fermi-LAT measurements strongly constrain DM decays into hadronic final states due to large production of gamma-rays as primary and secondary particles. On the other hand, present data do not exclude the presence of a second component in the neutrino flux coming from leptophilic decaying DM particles, especially when the astrophysical power-law component is assumed to follow the prior deduced by through-going muon neutrinos~\cite{Chianese:2019kyl}.

While the expected neutrino energy spectrum from astrophysical sources and DM particles might be similar, they exhibit very different angular distributions for the arrival directions of neutrino events. In particular, astrophysical sources are expected to be uniformly distributed in the full sky, while DM neutrino signal has some level of correlation with the galactic center where a high density of DM particles is typically expected. Such a difference in the angular distribution is therefore a key feature that would allow one to firmly discriminate between astrophysical and DM neutrino fluxes. Different analyses have pointed out that a galactic contribution consistent with gamma-rays can only account for less than 10\% of the observed neutrino flux~\cite{Ahlers:2015moa,Denton:2017csz,Aartsen:2017ujz} (see also Ref.~\cite{Chianese:2016opp}). These studies assume the galactic contribution to correlate with the galactic bulge and disc, as expected for astrophysical galactic sources. On the other hand, angular studies based on a more extended template for the galactic contribution have instead shown a preference for a galactic component in the neutrino flux~\cite{Esmaili:2014rma,Chianese:2016opp,Neronov:2015osa,Neronov:2016bnp}. In this context, a DM signal would indeed contribute to the neutrino data as a quite extended galactic diffuse emission identified with the DM halo, as well as an extragalactic isotropic emission. These two galactic and extragalatic DM contributions can be of the same order of magnitude depending on the DM model considered. 

In this paper, we investigate the two-component hypothesis (astrophysical power-law plus a DM signal) by analyzing only the angular distribution of the simulated neutrino sky maps by using the angular power spectrum. We test our two-component hypothesis with Monte Carlo method for a wide range of values for the lifetime and cross-section of the Dark Matter models. We assume the astrophysical component to be fixed by the 10-year through-going muon neutrinos data~\cite{Stettner:2019tok}, while we consider different models for the DM halo profile and boost factor, which provide different angular distribution for the DM component. Under the assumption of an isotropic neutrino flux, we place a constraint on the level of anisotropy induced by a DM component, which is then translated on a bound for DM lifetime and cross-section in case of decaying and annihilating DM particles, respectively. The two-component model is therefore tested for various DM parameters with 6-year of HESE observations, as well as for future neutrino data with IceCube-Gen2~\cite{Ackermann:2017pja} and KM3NeT~\cite{Adrian-Martinez:2016fdl} exposures. While the former will collect many more neutrino events thanks to the larger volume, the latter is expected to be more sensitive to the galactic center and, consequently, to a potential DM component. Our current and future constraints are weaker than the ones reported in other analyses, but they are more robust since they rely only on the angular distribution of neutrinos. However, we show that, after 10 years of observation, IceCube-Gen2 and KM3NeT will firmly probe the current best-fit for the DM contribution of the two-component flux reported in Ref.~\cite{Chianese:2019kyl}, thanks to the expected clustering of events towards the galactic center.

The paper is organized as follows. In Sec.~\ref{sec:flux}, we delineate the main features of a neutrino flux originated by astrophysical sources and by decaying and annihilating DM particles. Section~\ref{sec:sky_map} describes how the neutrino sky map is computed for the null hypothesis (atmospheric background plus an astrophysical power-law component) and for the signal hypothesis (atmospheric background, an astrophysical power-law and DM components) that we want to test. In Sec.~\ref{sec:aps}, we discuss the angular power spectrum analysis based on Monte Carlo simulations of the neutrino sky map. In Sec.~\ref{sec:results}, we report the current constraints and future sensitivity for DM lifetime and cross-section, while in Sec.~\ref{sec:discussion}, we discuss the dependence of the results on different astrophysical and DM models. Finally, in Sec.~\ref{sec:conclusions}, we draw our conclusions.

\section{Neutrino flux~\label{sec:flux}}

The neutrino flux observed with IceCube is typically explained in terms of the atmospheric neutrino background and an astrophysical power-law component. The former describes the neutrinos produced in the atmosphere through the interactions of cosmic rays, while the latter comes from a population of diffuse unresolved astrophysical sources. We consider only conventional atmospheric neutrinos, which are produced by decays of pions and kaons. Even though they are subdominant for $E_{\nu}>60~$TeV (lower energy threshold considered in this analysis), we include these in our sky maps by adopting the flux model from Ref.~\cite{Honda:2015fha, Feyereisen:2016fzb}. Furthermore, there might be a contribution coming from decaying or annihilating DM particles. In the following two subsections, we discuss the astrophysical and DM components, respectively.

\subsection{Astrophysical power-law component}

For the astrophysical component, we adopt an isotropic neutrino flux parameterized by a single power-law,  having the following form:
\begin{equation}
    \frac{d\Phi_{\nu_\alpha+\bar{\nu}_\alpha}^{\rm astro}}{dE_\nu\,d\Omega} =\Phi_0 \,\left(\frac{E_{\nu}}{100 \rm{TeV}}\right)^{-\gamma} \,,
\end{equation}
where $\Phi_0$ is the normalization of the astrophysical neutrino flux per neutrino flavor $\alpha$ at $100~$TeV (in units of $10^{-18} \, \rm{GeV}^{-1}\,\rm{cm}^{- 2}\,\rm{s}^{-1}\,\rm{sr}^{-1}$), and $\gamma$ is the spectral index. The parameters for the normalization and the spectral index are taken from the latest best-fits by IceCube for the two data samples of through-going~\cite{Stettner:2019tok} and HESE~\cite{Schneider:2019ayi} events. In particular, we have $\gamma^{\rm TG} = 2.28\pm^{+0.08}_{-0.09}$ and $\gamma^{\rm HESE} = 2.89^{+0.20}_{-0.19}$, and $\Phi_0^{\rm TG} = 1.44^{+0.25}_{-0.24}$ and $\Phi_0^{\rm HESE} = 2.15^{+0.49}_{-0.15}$.

\subsection{Dark matter signal}

High-energy neutrinos can be produced by DM particles through their decays or annihilations. The corresponding differential neutrino flux is the sum of the galactic (gal.) contribution from the Milky-Way halo and the extragalactic (ext.gal.) component:
\begin{equation}
\frac{d\Phi_{\nu_\alpha+\bar{\nu}_\alpha}^{\rm DM}}{dE_{\nu}\,d\Omega} = \sum_{\beta} P_{\alpha\beta} \left[ \frac{d\Phi_{\nu_\beta+\bar{\nu}_\beta}^{\rm gal.}}{dE_{\nu}\,d\Omega} + \frac{d\Phi_{\nu_\beta+\bar{\nu}_\beta}^{\rm ext. gal.}}{dE_{\nu}\,d\Omega} \right]\,,
\label{eq:DM_flux}
\end{equation}
where the quantities $P_{\alpha\beta}$ take into account the neutrino flavour oscillations during the propagation from the source to the Earth. Using the mixing angles obtained by the recent global neutrino fit~\cite{Capozzi:2018ubv,Capozzi:2020qhw}, we get
\begin{equation}
\begin{array}{lclcl}
P_{ee}=0.551\,, & \qquad & P_{e\mu}=0.191\,, & \qquad & P_{e\tau} = 0.258\,, \\
&&&&\\
P_{\mu\mu}=0.427\,, & \qquad & P_{\mu\tau}=0.383\,, & \qquad & P_{\tau\tau} = 0.359\,.
\end{array}
\end{equation}
The two contributions in Eq.~\eqref{eq:DM_flux} take different expressions in case of a decaying (dec.) or annihilating (ann.) DM signal. For the decaying scenario, we have
\begin{eqnarray}
\left. \frac{d \Phi_{\nu_\beta+\bar{\nu}_\beta}^{\rm gal.}}{dE_\nu d\Omega} \right|_{\rm dec.} & = & \frac{1}{4\pi \, m_{\rm DM} \, \tau_{\rm DM}} \frac{d N_\beta}{d E_\nu} \int_0^\infty ds \, \rho\left[r\left(s,\ell,b\right)\right]\,,\\
\left. \frac{d \Phi_{\nu_\beta+\bar{\nu}_\beta}^{\rm ext.gal.}}{dE_\nu d\Omega} \right|_{\rm dec.} & = &  \frac{\Omega_{\rm DM}\rho_c}{4\pi \, m_{\rm DM} \, \tau_{\rm DM}} \int_0^\infty dz \,  \frac{1}{H\left(z\right)}\left.\frac{d N_\beta}{d E'_\nu}\right|_{E'_\nu=E_\nu\left(1+z\right)}\,,
\end{eqnarray}
where $m_{\rm DM}$ and $\tau_{\rm DM}$ are the mass and the lifetime of DM particles, respectively. The quantity $\rho\left(r\right)$ is the DM halo density profile of the Milky Way that is integrated over the line-of-sight $s$. In particular, the radial distance is $r = \sqrt{s^2+R^2-2sR\cos\ell\cos b}$ with $R = 8.5$~kpc and $\left(b,\ell\right)$ are the galactic coordinates. The extragalactic component is instead proportional to the integral over the redshift $z$ where $\rho_c = 5.5 \times 10^{-6}\,{\rm GeV \, cm}^{-3}$ is the critical energy density and  $H\left(z\right)$ is the Hubble expansion rate as measured by Planck~\cite{Ade:2015xua}. Lastly, both fluxes depend on the energy spectrum ${\rm d}N_\beta/{\rm d}E_\nu$ of $\beta$-flavor neutrinos produced by DM particles. This quantity is obtained by taking the spectra from the \texttt{PPPC} package~\cite{Cirelli:2010xx} and following the scaling procedure for DM masses larger than 100~TeV reported in Ref.~\cite{Chianese:2016kpu}. As discussed in Ref.~\cite{Chianese:2019kyl}, such a procedure provides energy spectra compatible with the ones computed using \texttt{PYTHIA}~\cite{Sjostrand:2014zea}.

In the case of annihilating DM, the galactic and extragalactic components are given by
\begin{eqnarray}
\left. \frac{d \Phi_{\nu_\beta+\bar{\nu}_\beta}^{\rm gal.}}{dE_\nu d\Omega} \right|_{\rm ann.} & = & \frac12 \frac{\left< \sigma v \right>}{4\pi \, m_{\rm DM}^2} \frac{{\rm d}N_\beta}{{\rm d}E_\nu} \int_0^\infty ds \, \rho^2\left[r\left(s,\ell,b\right)\right]\,,\\
\left. \frac{d \Phi_{\nu_\beta+\bar{\nu}_\beta}^{\rm ext.gal.}}{dE_\nu d\Omega} \right|_{\rm ann.} & = & \frac12 \frac{\left< \sigma v \right> \, \left(\Omega_{\rm DM}\rho_c\right)^2}{4\pi \, m_{\rm DM}^2} \int_0^\infty dz \, \frac{B\left(z\right) \, \left(1+z\right)^3}{H\left(z\right)}\left.\frac{{\rm d}N_\beta}{{\rm d}E'_\nu}\right|_{E'_\nu=E_\nu\left(1+z\right)}\,,
\end{eqnarray}
respectively.
In this case, the neutrino flux is instead proportional to the DM thermally averaged cross-section $\left< \sigma v \right>$. Moreover, the redshift integral contains the boost factor (or clumpiness factor) $B\left(z\right)$, which accommodates the impact of the amount of DM substructures in the intergalactic medium (see Ref.~\cite{Ando:2019xlm} for a recent review) as well as the halo mass function~\cite{Sheth:1999mn, Sheth:1999su}. Here, we do not consider the effect of the presence of DM subhalos in our galaxy (see for example Ref.~\cite{Zavala:2014dla}).

\begin{figure}
    \centering
    \includegraphics[scale=.45]{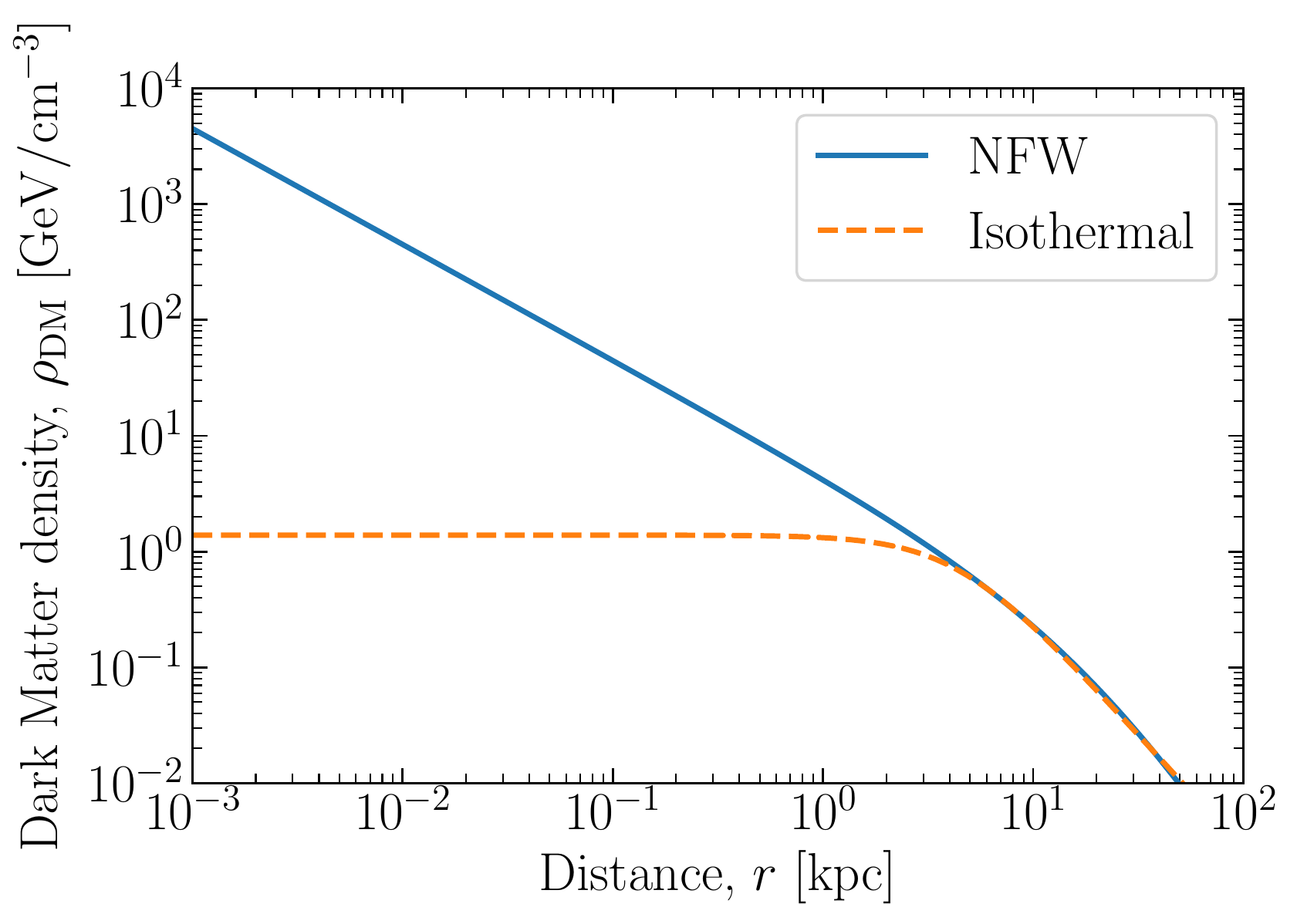}
    \hskip7.mm
    \includegraphics[scale=.43]{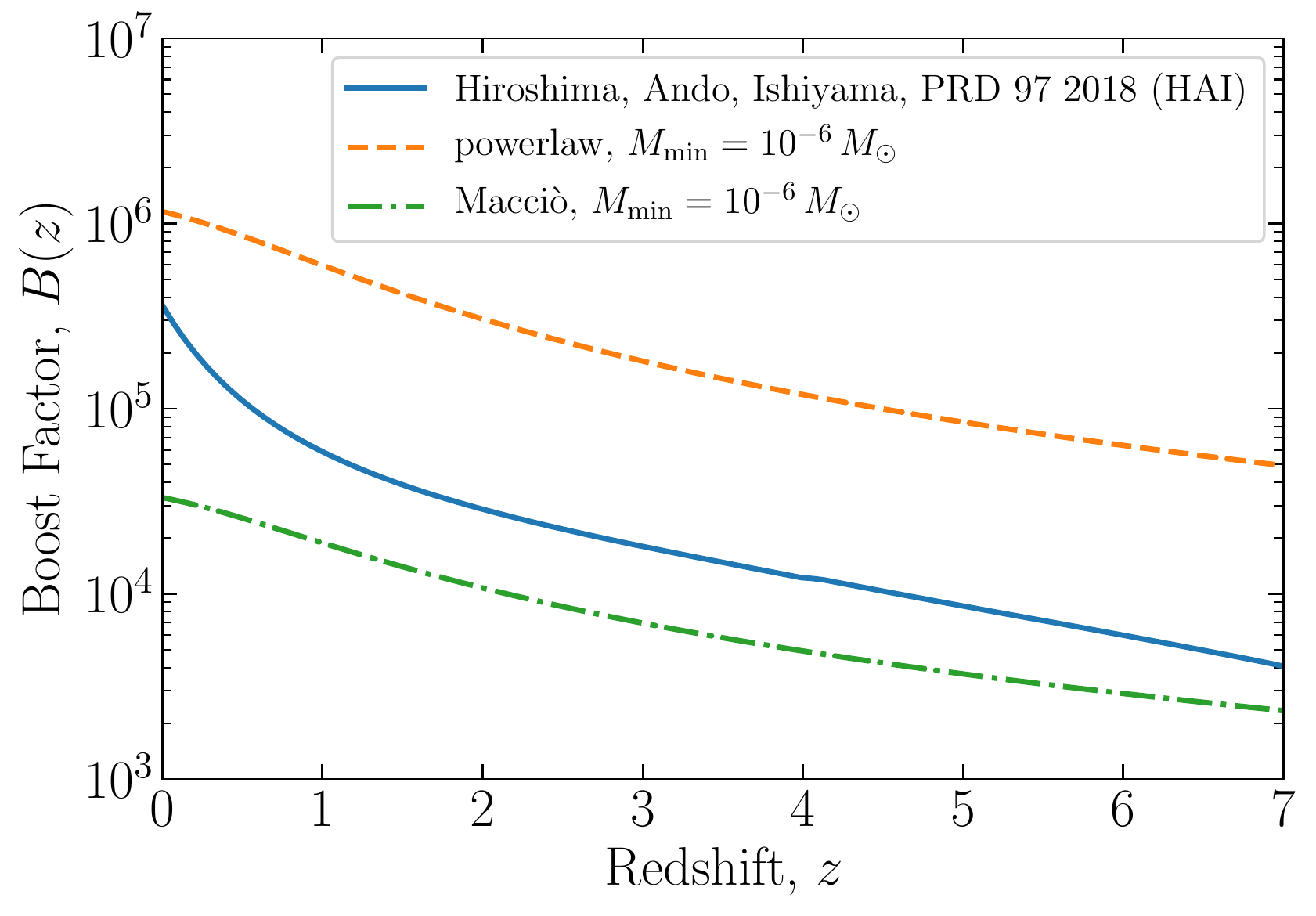}
\caption{{\bf Different dark matter models for halo density profile and boost factor considered in this paper.} {\it Left}: DM halo density profile as a function of the distance from the galactic center for the NFW and isothermal models. {\it Right}: DM boost factor as a function of the redshift for different models.}
    \label{fig:DM_properties}
\end{figure}

In the following, we consider two different final particle states of DM decay/annihilation providing very different neutrino energy spectra: the leptonic channel into $\tau$ leptons and the handronic one into top quarks as representative of hard and soft neutrino spectra, respectively. Most importantly, the DM signal depends on the halo density profile and the boost factor, which weigh the galactic and the extragalactic (isotropic) components, respectively. Different choices of these quantities provide very different levels of anisotropy. We analyze two DM halo density profiles that provide very different angular correlation with the galactic center: the Navarro-Frenk-White distribution (NFW)~\cite{Navarro:1995iw} and the Isothermal profile (ISO)~\cite{Begeman:1991iy}. The former leads to large neutrino fluxes towards the galactic center, while the latter is essentially uniform. The two profiles are respectively given by~\cite{Cirelli:2010xx}
\begin{equation}
\rho^{\rm NFW} \left(r\right)= \frac{\rho_s}{r/r_s\left(1+r/r_s\right)^2}\,,
\label{eq:NFW}
\end{equation}
with $r_s = 24$~kpc and $\rho_s = 0.18 \, {\rm GeV \, cm}^{-3}$, and
\begin{equation}
\rho^{\rm ISO} \left(r\right)= \frac{\rho_s}{1+\left(r/r_s\right)^2}\,,
\label{eq:ISO}
\end{equation}
with $r_s = 4.4$~kpc and $\rho_s = 1.9 \,{\rm GeV \, cm}^{-3}$. The two halo density profiles as a function of the distance from the galactic center are reported in the left panel of Fig.~\ref{fig:DM_properties}. In the case of DM annihilation, in order to accommodate uncertainties on the DM substructures, we examine three different models for the cosmological boost factor $B(z)$: the semi-analytical model described in Ref.~\cite{Hiroshima:2018kfv} (hereafter dubbed as HAI), the ``Macci{\`o}'' model~\cite{Maccio:2008pcd} that is based on an earlier phenomenological model~\cite{Bullock:1999he}, and ``power-law'' model~\cite{Neto:2007vq,Maccio:2008pcd} that is obtained by extrapolating the results for the concentration parameter from ab-initio N-body simulations, by taking a minimum halo mass of $10^{-6}M_\odot$~\cite{Cirelli:2010xx}. We note that the simple power-law extrapolation of the concentration-mass relation, which yields significant overestimate of the neutrino flux, is no longer considered realistic. The right panel of Fig.~\ref{fig:DM_properties} shows the boost factor as a function of the redshift $z$ for the three models. It is worth noticing that the HAI model is valid until redshift $z \simeq 7$ and we simply take $B_{\rm HAI}(z) = 1$ for $z\gtrsim7$. It is worth observing that regions at higher redshift provide in general a subdominant contribution to the total DM component. In particular, due to the redshift of the neutrino energy, the regions for $z\gtrsim7$ would marginally contribute to the neutrino spectrum at neutrino energies $E_\nu \lesssim \mathcal{O}(100~\mathrm{TeV})$ only in case of dark matter masses $m_\mathrm{DM} \gtrsim 1~\mathrm{PeV}$. We have checked that for all the different boost factor models this contribution is typically subdominant when analyzing the neutrino sky above 60~TeV, which is the lower energy threshold in the present analysis (see the discussion below).

\section{Sky map \label{sec:sky_map}}

We simulate the neutrino sky map above 60~TeV under two different assumptions:
\begin{itemize}
    \item {\bf Null hypothesis:} in addition to the conventional atmospheric neutrino background (atm), we consider an isotropic astrophysical component only
\begin{equation}
    \frac{d\Phi^{\rm null}}{dE_{\nu}d\Omega} = \frac{d\Phi_{\rm{atm}}}{dE_{\nu}d\Omega} +  \frac{d\Phi_{\rm{astro}}^{\rm{HESE}}}{dE_{\nu}d\Omega} \label{eq:null}\,,
\end{equation}
where we use the HESE best-fit for the astrophysical component.
    \item {\bf Signal hypothesis:} we include a DM component so having
    \begin{equation}
    \frac{d\Phi^{\rm signal}}{dE_{\nu}d\Omega} = \frac{d\Phi_{\rm{atm}}}{dE_{\nu}d\Omega} +  \frac{d\Phi_{\rm{astro}}^{\rm{TG}}}{dE_{\nu}d\Omega} + \frac{d\Phi_{\rm{DM}}}{dE_{\nu} d\Omega} \label{eq:signal}\,.
    \end{equation}
    In this case, the astrophysical power-law is fixed by the TG data sample. Hence, the neutrino flux originated by DM is the additional component required to alleviate the tension between HESE and TG data sets. Then, we test various parameters for decaying and annihilating DM signals.
\end{itemize}

For each flux component, the expected number of observed neutrino events in a region of the sky $\Delta \Omega$ identified by the position $\theta$ (declination) and $\phi$ (right ascension) is derived as follows
\begin{equation} \label{eq:Number}
N_{\nu} \left(\theta, \phi \right)  = \int_{\Delta \Omega} \,d\Omega \, \int_{E_{\rm th}}^{E_{\rm max}} \, dE_{\nu} \, \sum_\alpha f_\alpha\,\frac{d\Phi_{\nu_\alpha+\bar{\nu}_\alpha}}{dE_{\nu}\,d\Omega} \, \mathcal{E}_\alpha(E_{\nu}, \Omega) \, \rm{vis}(\Omega)\,.
\end{equation}
Here, $\mathcal{E}_\alpha(E_{\nu}, \Omega) = T_{\rm obs}\,A_{\rm eff, \alpha}(E_{\nu}, \Omega)$ is the detector's exposure for neutrino flavor $\alpha$ with $T_{\rm obs}$ $A_{\rm eff, \alpha}$ being the observation time and the detector's effective area, respectively. The quantity $\rm{vis}(\Omega)$ is the visibility function quantifying the fraction of the year during which a point in the sky can be observed by the telescope. In general, it is a non-trivial function that depends on the Earth rotation and the veto technique of the telescope to suppress the atmospheric background. In the case of IceCube, we consider the HESE effective area~\cite{Aartsen:2013jdh} and $\rm{vis}^{\rm IC}(\Omega) = 1$ thanks to the muon self-veto~\cite{Gaisser:2014bja,Arguelles:2018awr}. The same is assumed for IceCube-Gen2 for which we take its effective area to simply be ten times larger than the HESE one. For KM3NeT, we use the exposure and the visibility function reported in Ref.~\cite{Adrian-Martinez:2016fdl}, and account for the Earth's absorption using the code \texttt{$\nu$FATE}~\cite{Vincent:2017svp}. In Eq.~\eqref{eq:Number}, the factor $f_\alpha$ takes into account the fraction of neutrinos producing an event with shower or track topology, so that $f^{\rm shower}_\alpha + f^{\rm track}_\alpha = 1$. For the sake of simplicity, we assume that electron and tau neutrinos produce only shower events, and therefore $f^{\rm shower}_\tau \simeq f^{\rm shower}_e \simeq 1$. On the other hand, only a fraction $f^{\rm track}_\nu = 0.8 $ of muon neutrinos provides track-like events, according to the probability of having charged current interaction, i.e. $\sigma_{CC} / (\sigma_{NC}+\sigma_{CC}) \simeq 0.8$~\cite{Gandhi:1998ri}. In order to reduce the background contamination (penetrating atmospheric muons and conventional atmospheric neutrinos) in IceCube, we consider only shower-like events. This implies that $f^{\rm IC}_\tau=f^{\rm IC}_e=1$ and $f^{\rm IC}_\mu=0.2$. In the case of KM3NeT, we instead examine only through-going track events for which $f^{\rm KM3NeT}_\tau=f^{\rm KM3NeT}_e=0$ and $f^{\rm KM3NeT}_\mu = 0.8$. Furthermore, for both telescopes, we set a lower energy threshold of $E_{\rm th}=60~$TeV in order to further suppress the background events, and an upper limit corresponding to the DM mass with $E_{\rm max}=m_{\rm{DM}}/2$ for decaying DM and $E_{\rm max}=m_{\rm{DM}}$ for annihilating DM scenarios. Such an upper energy threshold corresponds to the maximum energy available for neutrinos produced by DM particles. It is worth noticing that these cuts on neutrino energies and the event topology selection for IceCube and KM3NeT considered in the present analysis are not aimed at providing the optimal strategy for data-analysis. Instead, we aim at highlighting the potentiality of analyzing the angular power spectrum of the neutrino sky with both the telescopes. Moreover, we stress that IceCube and KM3NeT are expected to provide different constraints on the dark matter parameters mainly due to the different sensitivity to the galactic center.

The simulated sky maps are therefore given by the sum of the following contributions:
\begin{equation}
    N_{\rm tot}\left(\theta, \phi \right) = N_{\rm atm} + N_{\rm astro} + N_{\rm DM} \,,
\end{equation}
where the DM events are further obtained by summing the galactic and extragalactic contributions
\begin{equation}
    N_{\rm DM} = N_{\rm DM}^{\rm gal.} + N_{\rm DM}^{\rm ext.gal.} \,.
\end{equation}
In the present analysis, we assume that the level of anisotropy in the neutrino flux is set only by the galactic DM component, while the other contributions are considered to be statistically isotropic. This is not true for the atmospheric flux measured by IceCube since the muon veto of the experiment highly suppresses the down-going atmospheric neutrinos. However, in our settings, above a neutrino energy of $60$~TeV, we expect less than 1 atmospheric neutrino event per year with a shower-like topology. This explains why we consider only shower events when analyzing IceCube. For KM3NeT, since the contamination of the atmospheric neutrino background above 60~TeV is still expected to be small, the null hypothesis flux can be statistically approximated to be isotropic.
\begin{figure}[t!]
\centering
\includegraphics[width=.32\textwidth]{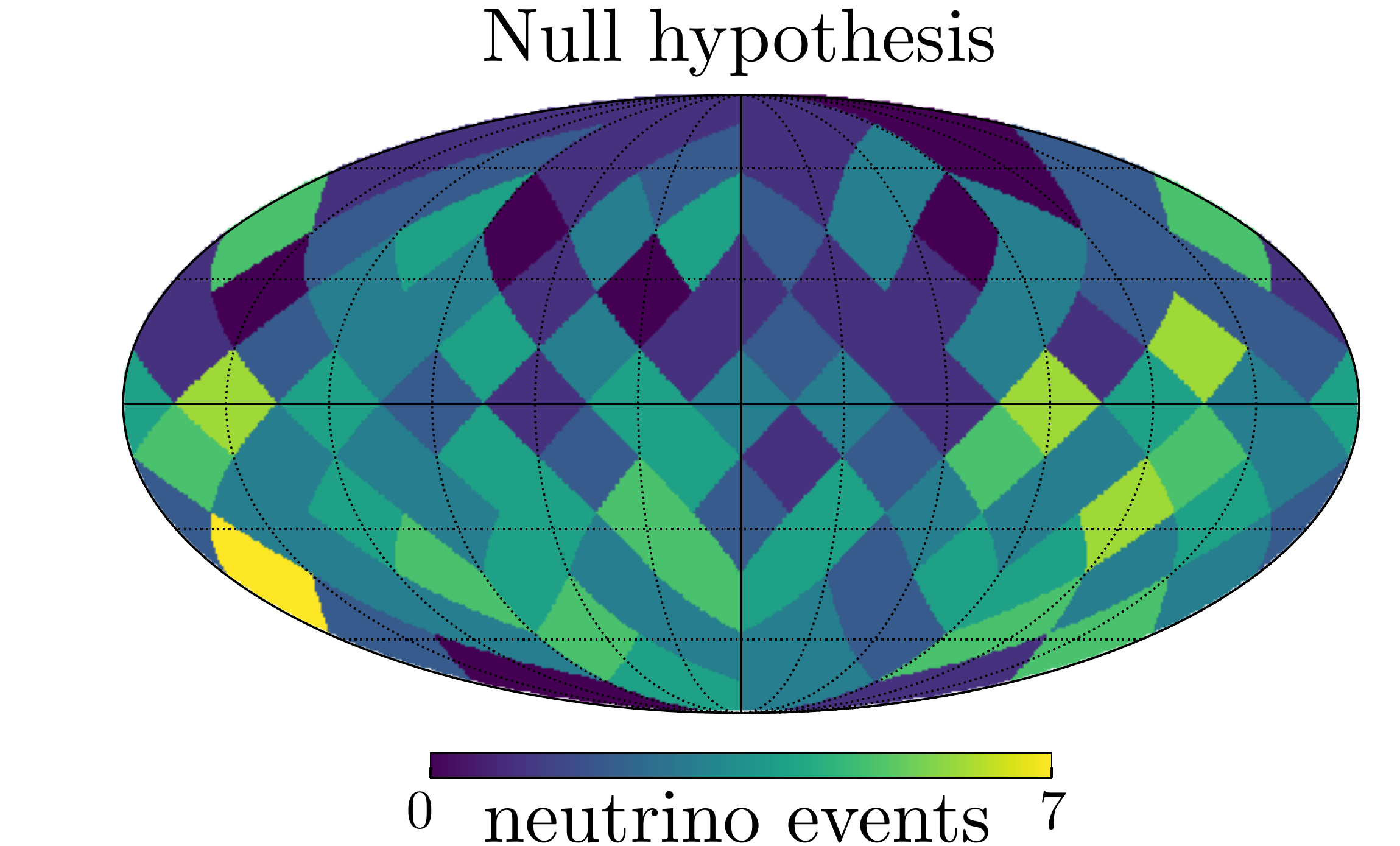}\hfill
\includegraphics[width=.32\textwidth]{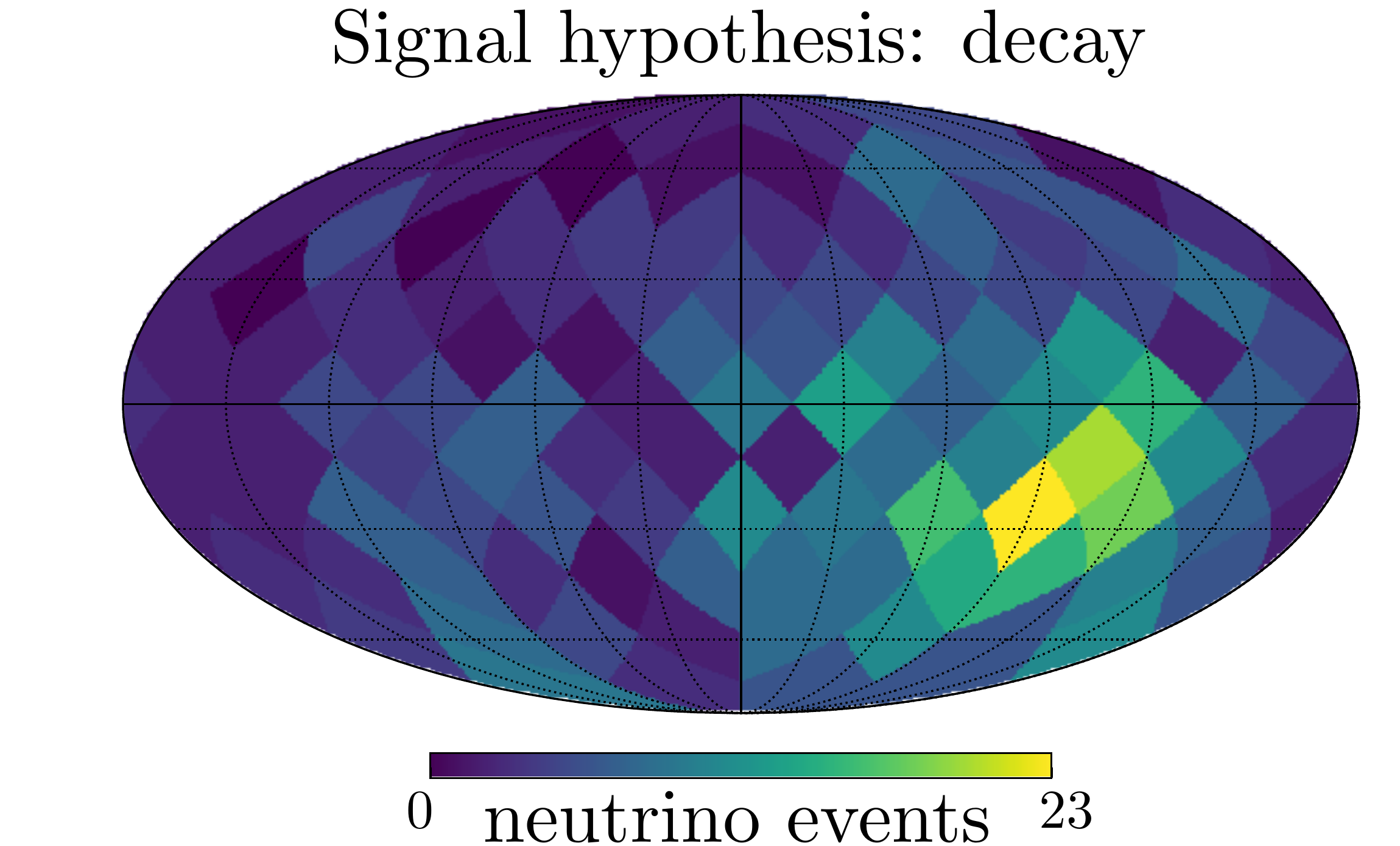}\hfill
\includegraphics[width=.32\textwidth]{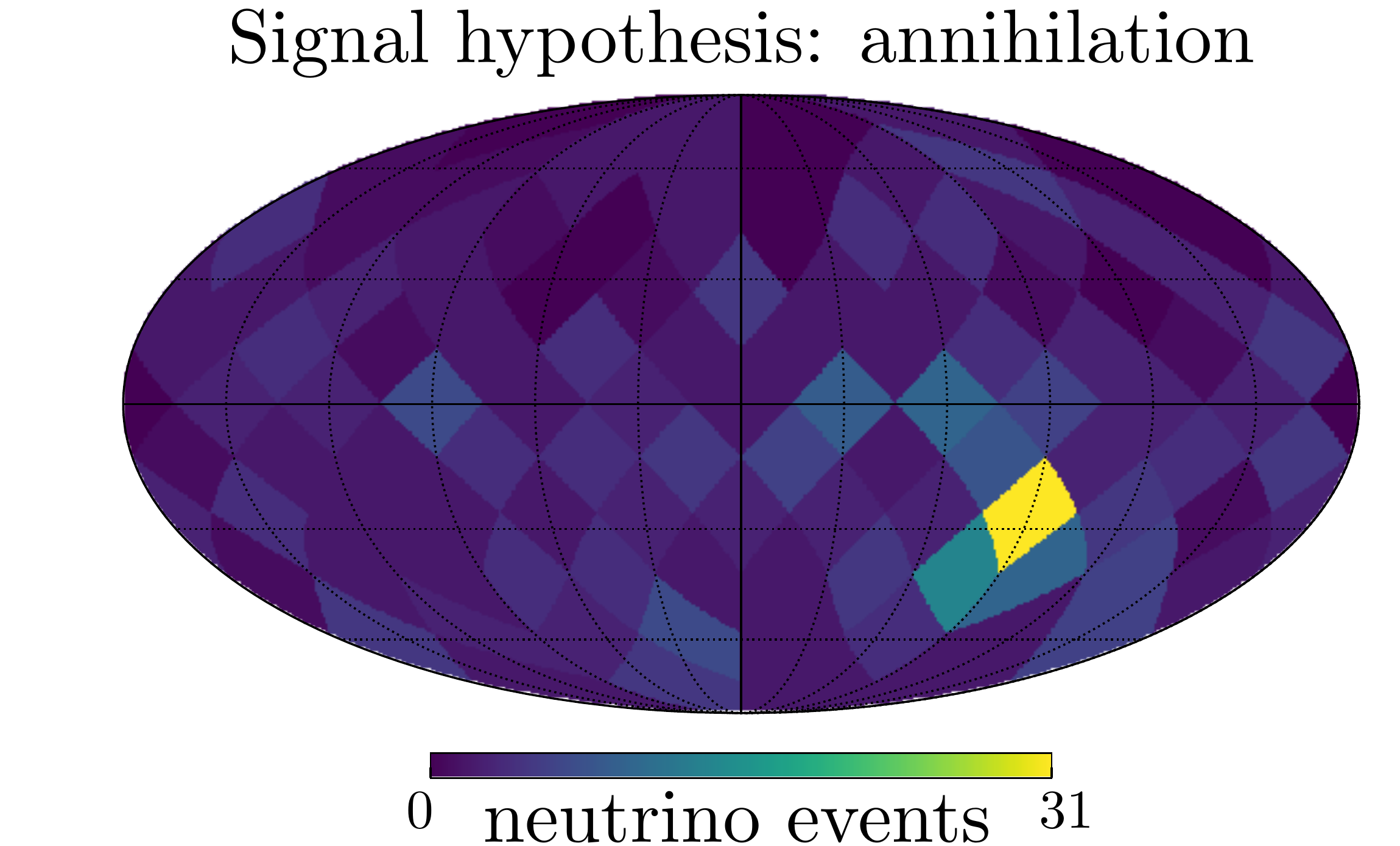}
\caption{{\bf Neutrino sky maps under different flux hypotheses after 10 years of observation in IceCube-Gen2.} The left panel shows the angular distribution of neutrino events under the null hypothesis of a nearly isotropic flux (atmospheric and astrophysical power-law components). The middle and right panels display the sky maps for the signal hypothesis with an additional component coming from decaying and annihilating DM particles, respectively. }
\label{fig:Skymaps}
\end{figure}

Hence, the ratio between anisotropic and isotropic number of neutrino events depends on DM parameters like for instance the density profile, boost factor and DM lifetime or cross-section. We thus aim to constrain DM parameters by testing our two-component model (signal hypothesis) with respect to an isotropic astrophysical model (null hypothesis). Figure~\ref{fig:Skymaps} illustrates the differences in angular patterns for the null hypothesis (left) and for two signal hypotheses with decaying (middle) and annihilating (right) DM component. These sky maps corresponds to one Monte Carlo realization with 10 years of IceCube-Gen2 exposure. The number of galactic neutrino events coming from annihilating DM is related to the halo density as $N^{\rm gal.}_{\rm DM}\propto\rho^2$, while $N^{\rm gal.}_{\rm DM}\propto\rho$ for decaying DM. This produces different amount of anisotropy as can be seen in the figure. In both cases, we have considered the NFW halo density and the semi-analytical HAI model for the boost factor. We will be examining these angular patterns with an angular power spectrum analysis, as described below. 

\section{Angular Power Spectrum analysis \label{sec:aps}}

The angular power spectrum (APS) shows to be a powerful probe to asses anisotropies on the neutrino sky~\cite{Dekker:2018cqu}. The fluctuations on the neutrino sky map is found by expanding the map into spherical harmonics:
\begin{equation}
N\left(\theta, \phi \right) = \sum_{\ell m} a_{\ell m} Y_{\ell m}\left(\theta, \phi \right) \,,
\end{equation}
where $N\left(\theta, \phi \right)$ is the neutrino event at declination ($\theta$) and right ascension ($\phi$), and $Y_{\ell m}(\theta, \phi)$ are the spherical harmonic functions. The APS is then given by averaging the expansion coefficients over the sky:
\begin{equation}
C_{\ell}' = \frac{1}{2\ell +1} \sum_{m=-\ell}^{\ell} |a_{\ell m}|^2 \,.
\end{equation}
We compute the APS using the numerical function \texttt{anafast} from the software package \texttt{HEALPix}~\cite{Gorski:2004by}. Since we are only interested in anisotropic effects, we want to remove any information on the number of events. We therefore remove the monopole and normalize the remaining coefficients as 
\begin{equation}
C_{\ell} = \frac{C_{\ell}'}{N^{2}_{\rm tot}} \qquad {\rm for}\qquad \ell>0 \,,
\end{equation}
where $N_{\rm tot}$ is the total number of neutrino events.
The first multipole moments show larger angular power due to the anisotropy of events coming from the galactic center, which is detectable on top of the isotropic distribution. We calculate the APS up to $\ell_\mathrm{max}=9$ for IceCube, IceCube-Gen2 and KM3NeT maps, which corresponds to the typical angular resolution of IceCube for HESE events, $11^{\circ}$~\cite{Aartsen:2013jdh,Aartsen:2014gkd}. Even though KM3NeT has a better angular resolution of $0.07^{\circ}$, including larger multipole moments will not improve our analysis since only the first multipole moments are of interest for our work. Indeed, the multipole moments with $\ell > 9$ related to the smooth and extended dark matter galactic component are not substantially different from the ones corresponding to an isotropic sky.

We perform Monte Carlo simulations for decaying and annihilating dark matter models and range the lifetime between $\tau_{\rm DM} =[10^{26},10^{31}]\,\rm{s}$ in the case of decaying dark matter and the cross-section between $\left<\sigma v\right>=[10^{-26},10^{-20}]\,\rm{cm^3 s^{-1}}$ for annihilating dark matter, both in steps of $\log(0.2)$. Additionally, we simulate purely isotropic astrophysical + atmospheric sky maps for the null hypothesis. Each case is simulated $10^5$ times for IceCube, IceCube-Gen2 and KM3NeT exposures and we calculate their APS. In order to have a statistical measure for the goodness of the models, we apply the following $\chi^{2}$,
\begin{equation}
\chi^{2} \left(C_{\ell}\right) = \sum_{\ell \ell'} \left(C_{\ell} - C_{\ell}^{\rm{mean}}\right) (\rm{Cov}_{\ell\ell'})^{-1} \left(C_{\ell'} - C_{\ell'}^{\rm{mean}}\right) \,,
\label{eq:chi2}
\end{equation}
where $C_{\ell}$ is the APS of one simulation, $C_{\ell}^{\rm{mean}}$ is the mean value and $\rm{Cov}_{\ell\ell'}$ is the covariance matrix, where $C_{\ell}^{\rm{mean}}$ and $\rm{Cov}_{\ell\ell'}$ are obtained from a complete set of simulation of the signal hypothesis. For each characterization of the model, we calculate the probability density function (PDF) of $\chi^2$, $P(\chi^2|\Theta)$ with $\Theta$ the parameters of the DM component. We then compute $\chi_{\rm data}^2\equiv \chi^2(C_{\ell}^{\rm{data}})$ from the observed neutrino sky to obtain the probability of having the same or more extreme values of $\chi^{2}$ by the following $p$--value,
\begin{equation}
p = \int_{\chi^2_{\rm data}}^\infty d\chi^2 P(\chi^2|\Theta).
\end{equation}
Models are constrained at 95\% confidence level (CL), which is equivalent to $p \leq 0.05$. For the forecast we use the isotropic best-fit model from Ref.~\cite{Schneider:2019ayi} (null hypothesis) as data for $\chi_{\rm data}^2\equiv \chi^2(C_{\ell}^{\rm{data}})$, which we simulate $10^5$ times as well. Instead of a single $p$--value, we therefore get a distribution of $10^5$ $p$--values, from which we consider the 68\% and 95\% contour bands symmetrically distributed around the mean.

\section{Results \label{sec:results}}
\begin{figure}[t!]
\centering
\includegraphics[width=.32\textwidth]{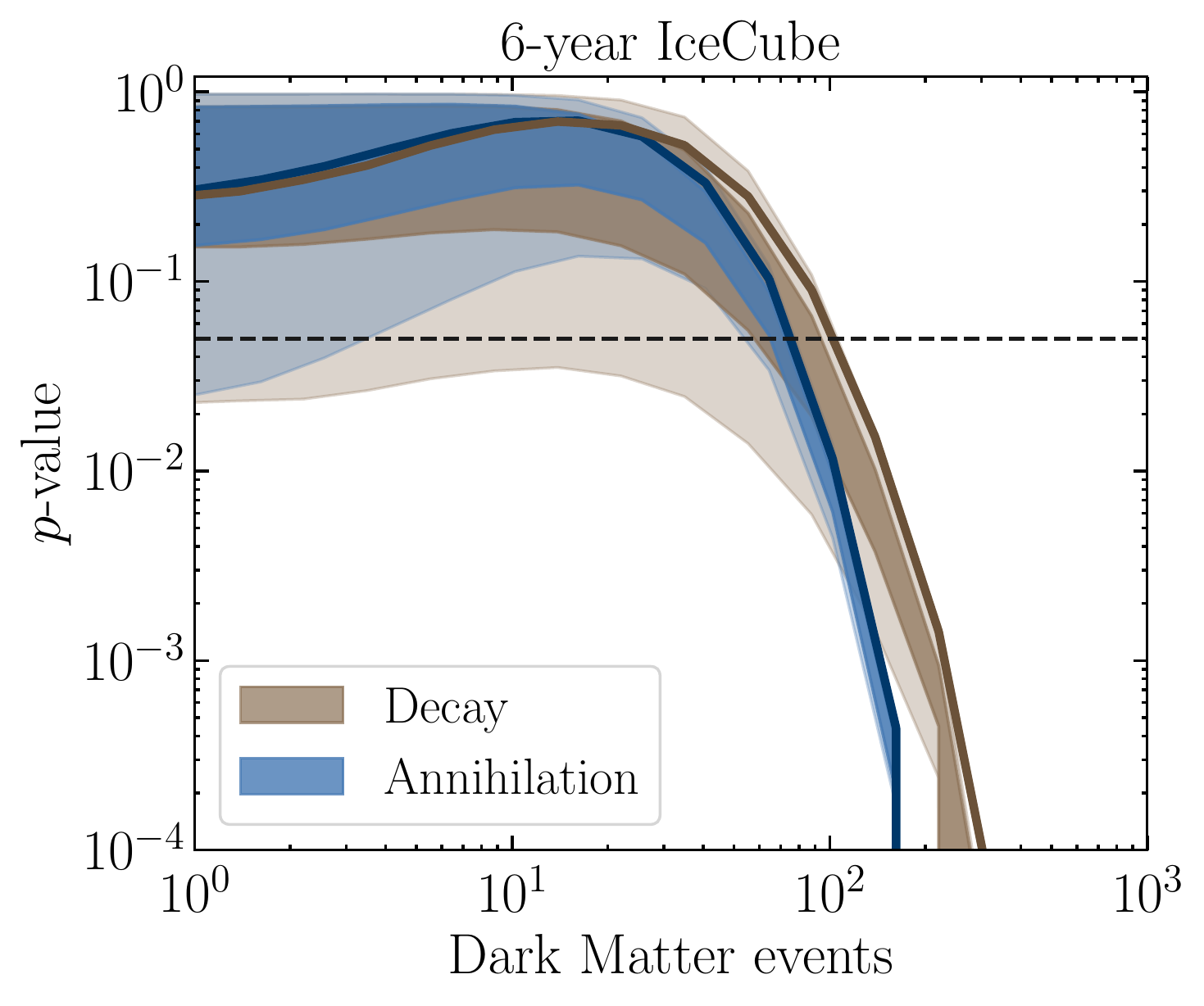}\hfill
\includegraphics[width=.32\textwidth]{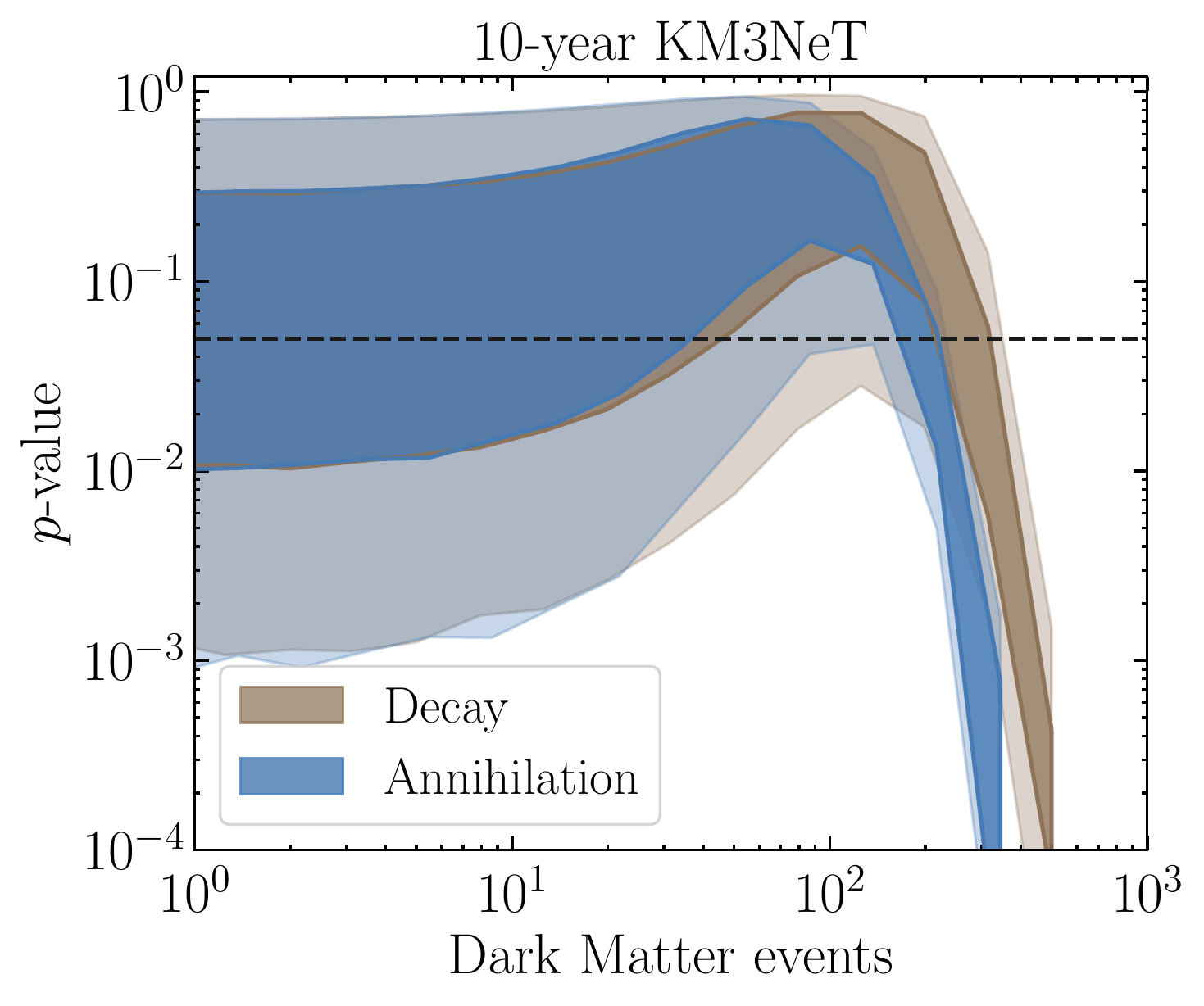}\hfill
\includegraphics[width=.32\textwidth]{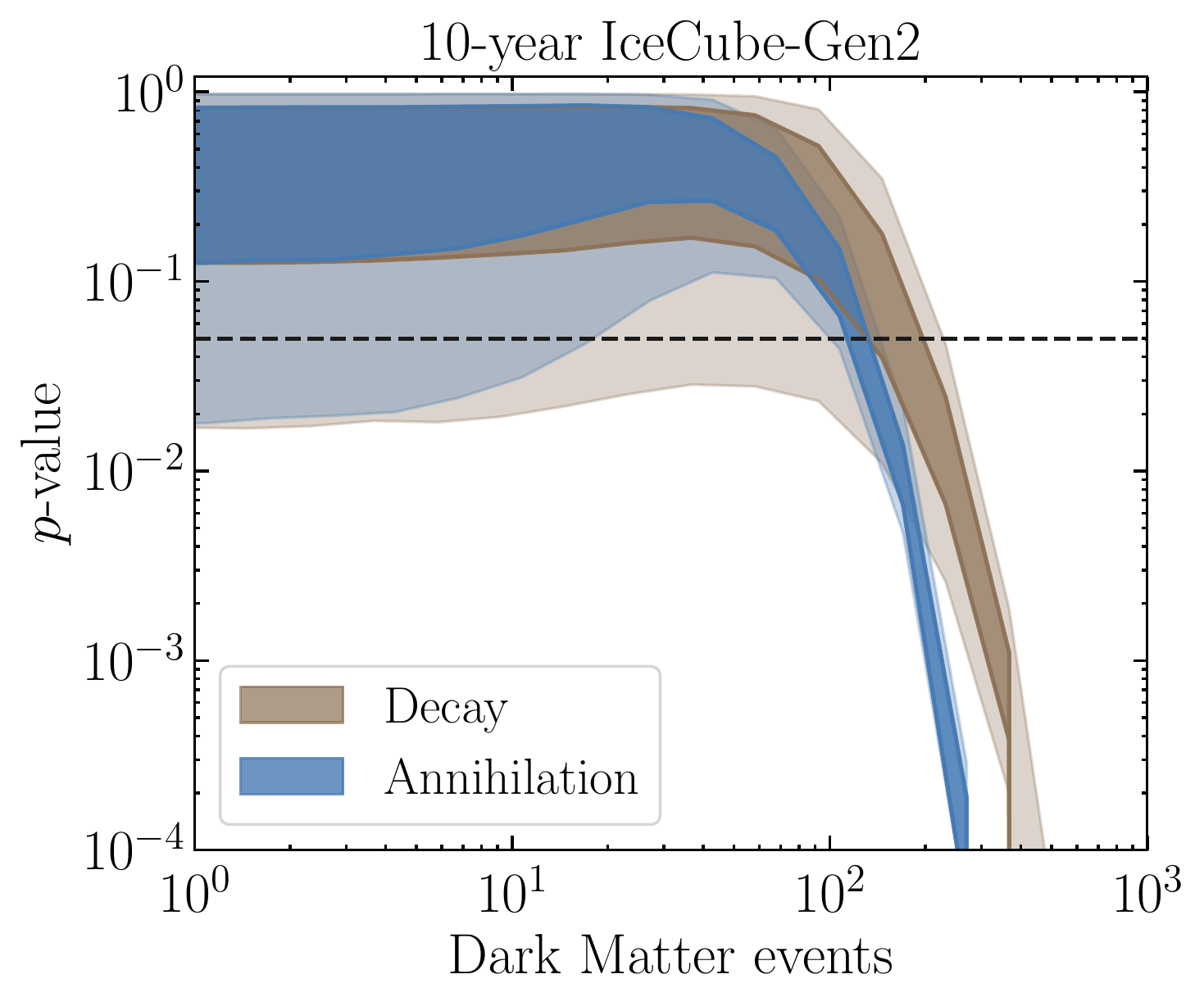}
\caption{{\bf Results from simulated sky maps for different observations.} The $p$--value bands as a function of the expected total DM neutrino events, for the signal hypothesis with annihilating (blue) and decaying (brown) DM particles into $\tau^+\tau^-$, assuming NFW halo profile and HAI boost factor. The DM mass is taken to be 200 and 400~TeV, respectively. The panels from left to right correspond to 6-year IceCube, 10-year KM3NeT and 10-year IceCube-Gen2 exposures, respectively. The dark and light colors refer to the 68\% and 95\% Monte Carlo realizations of the null-hypothesis sky map. The horizontal dashed line highlights the exclusion limit of $p=0.05$. In the left panel, the solid lines show the constraints from current  6-years IceCube HESE data.}
\label{fig:pval}
\end{figure}

We analyze the APS of the simulated sky maps obtained under the null hypothesis and the signal hypothesis for decaying and annihilating DM particles. In this section, we assume the NFW profile and the HAI boost factor as benchmark scenario. In Fig.~\ref{fig:pval} we report the results for current data with 6-year exposure time in IceCube (left panel), and future observations with 10 years of data-taking with KM3NeT (middle panel) and IceCube-Gen2 (right panel). The bands cover different Monte Carlo realizations of the null-hypothesis sky map used to test the DM scenarios of annihilation (blue) and decay (brown) into the leptonic final state of $\tau^+ \tau^-$ with a DM mass of 200 and 400~TeV, respectively. In particular, we show the 68\% (dark colors) and 95\% (light colors) contour bands from Monte Carlo simulations. In the left panel, we report the APS fits of the 6-year IceCube HESE observations, shown by the solid lines.\footnote{We perform the APS analysis on the 6-year IceCube HESE data because the angular coordinates of the events in the updated 7.5-year data-set are not yet public.} The 6-year HESE data consist of 33 neutrino events in the energy range 60--200~TeV, where 200~TeV corresponds to the maximum energy of neutrinos produced by DM particles. The horizontal black dashed line represents the exclusion limit of $p=0.05$ below which the two-component hypothesis is rejected. As can been seen in the plots, the APS analysis constrains the total number of neutrino events related to the DM signal: a larger DM contribution would produce a higher anisotropy towards the galactic center in contrast with the null hypothesis. On the other hand, when the number of DM neutrino events is small, the signal hypothesis becomes indistinguishable from the null hypothesis due to a negligible DM contribution. In this case, the distribution of $p$--values simply corresponds to the Monte Carlo Poisson noise of the $10^5$ null-hypothesis realizations. Such a behaviour can be seen in all the plots. In all the cases, when decreasing the number of DM neutrino events, the $p$--values move from being very small due to the strong disagreement between the null and the signal hypotheses to being dominated by the Monte Carlo Poisson noise.\footnote{Note that the null-hypothesis (Eq.~\eqref{eq:null} and the signal hypothesis (Eq.~\eqref{eq:signal}) are not nested, according to the different assumption on the astrophysical flux component.} Therefore, in order to provide conservative constraints and future sensitivity on the total DM events at 95\% CL ($p$-value smaller than 0.05), we consider the upper 95\% bound and the mean of the $p$-value distribution obtained by the Monte Carlo simulations of the null-hypothesis.

It is worth observing that the constraints on the annihilating scenario are stronger than the ones for decaying one. This is indeed expected due to the more concentrated contribution from the galactic halo in case of annihilating DM particles (see Fig.~\ref{fig:Skymaps}). However, as discussed in Sec.~\ref{sec:discussion}, the constraints on the total DM events depend on the models for the halo profile and the boost factor, since they set the relative contribution between galactic and extragalactic DM fluxes. On the other hand, the limits presented here are slightly dependent on the DM mass and almost independent of the annihilating/decaying channel. Indeed, the whole analysis is not very sensitive to the energy spectrum of neutrinos since the sky map is integrated over neutrino energy.
\begin{figure}[t!]
\centering
\includegraphics[width=.48\textwidth]{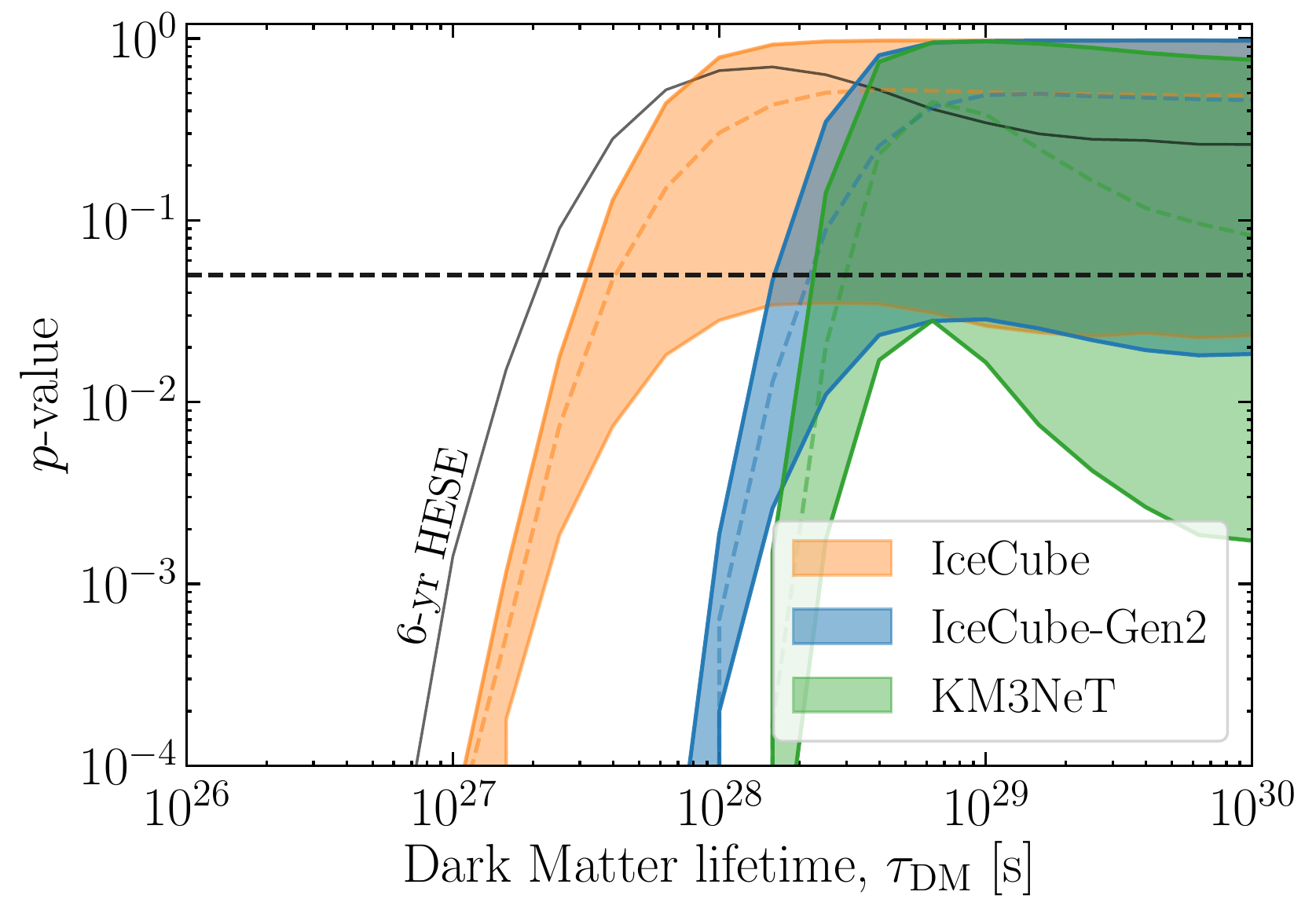}\hskip5.mm
\includegraphics[width=.48\textwidth]{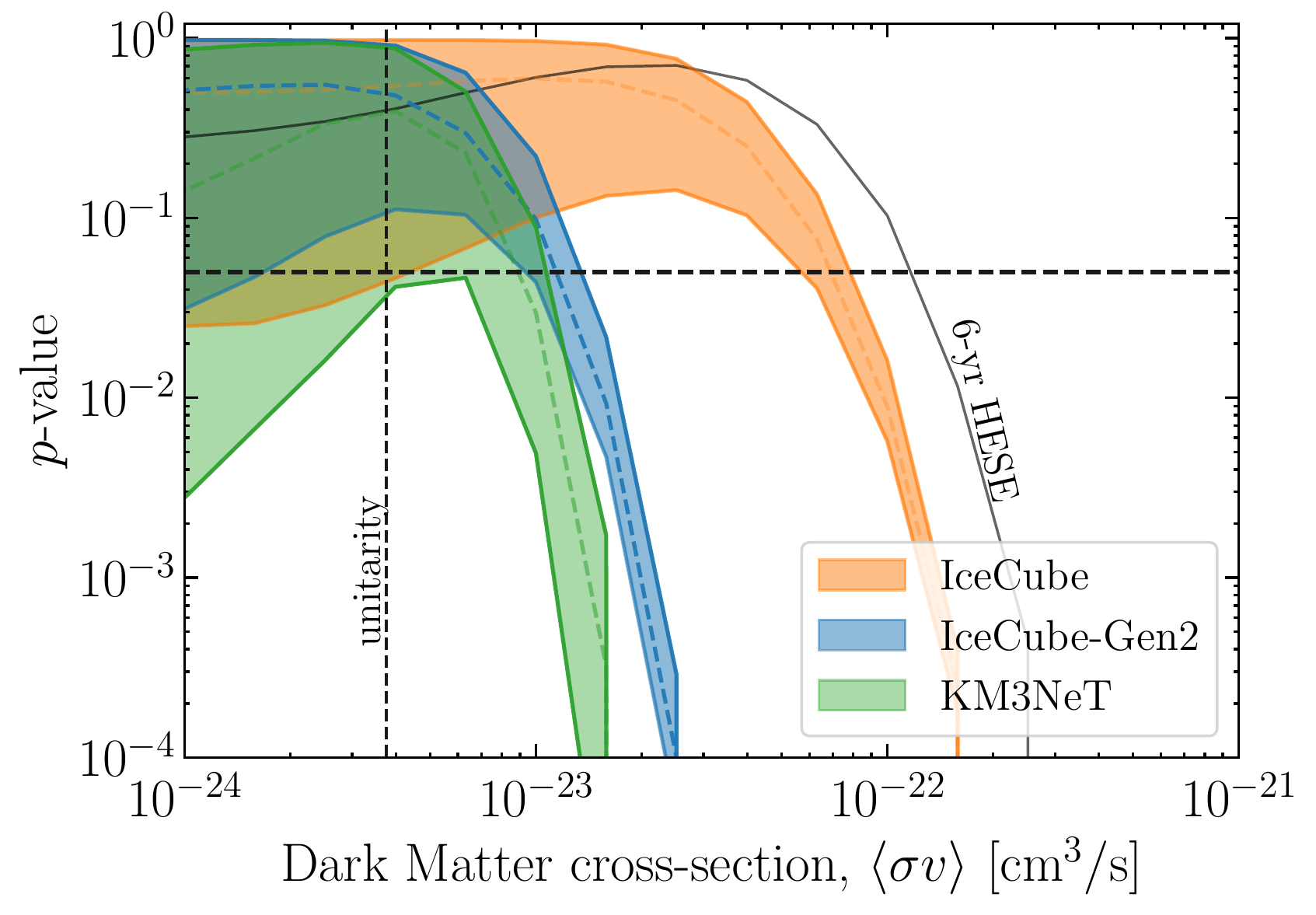}
\caption{\textbf{\textit{p}--values from simulated sky maps after 10-year observations with different neutrino telescopes.} The 95\% bands of $p$--values from Monte Carlo simulations as a function of DM lifetime (left) and cross-section (right) for $\tau^+\tau^-$ final state and DM mass of 400 and 200~TeV, respectively. The different bands represent 10-year observations with IceCube HESE (orange), IceCube-Gen2 HESE (blue) and KM3NeT through-going muon neutrinos (green) exposures. The dot-dashed lines inside the 95\% bands show the median $p$-values obtained from simulations. In the right panel, the vertical black solid line is the unitarity bound on cross-section.}
\label{fig:limits}
\end{figure}

The upper bounds on the total DM neutrino events are translated into constraints on the DM lifetime (left panel) and cross-section (right panel) in Fig.~\ref{fig:limits}. In this figure, we compare the sensitivity to DM parameters of IceCube (orange), KM3NeT (green) and IceCube-Gen2 (blue) after 10 years of observations. Remarkably, KM3NeT is found to be more sensitive to the DM models considered, so providing stronger constraints with respect to IceCube and IceCube-Gen2. This is indeed expected thanks to the KM3NeT geographical position that is more suitable to observe the galactic center via through-going neutrinos. On the other hand, the observation of the galactic center with IceCube and its future upgrade IceCube-Gen2 is limited by a smaller effective area due to the requirement of having down-going contained events. However, this is not enough to reach the unitarity limit (vertical solid line in Fig.~\ref{fig:limits}) on the DM cross-section implying $\langle \sigma v \rangle \lesssim 4\times 10^{-24}~\rm cm^3/s$ for $m_{\rm DM} = 200~\rm TeV$~\cite{Griest:1989wd}. Finally, in the figure, we also report the current constrains from the 6-year IceCube HESE data with the solid black lines. In particular, we have $\tau_{\rm} >  1.9\times10^{27}\,\rm s$ and $\langle \sigma v \rangle < 1.17\times10^{-22}\,\rm cm^3/s$ corresponding to $p$--values smaller than 0.05.
\begin{figure}[t!]
\centering
\includegraphics[width=.48\textwidth]{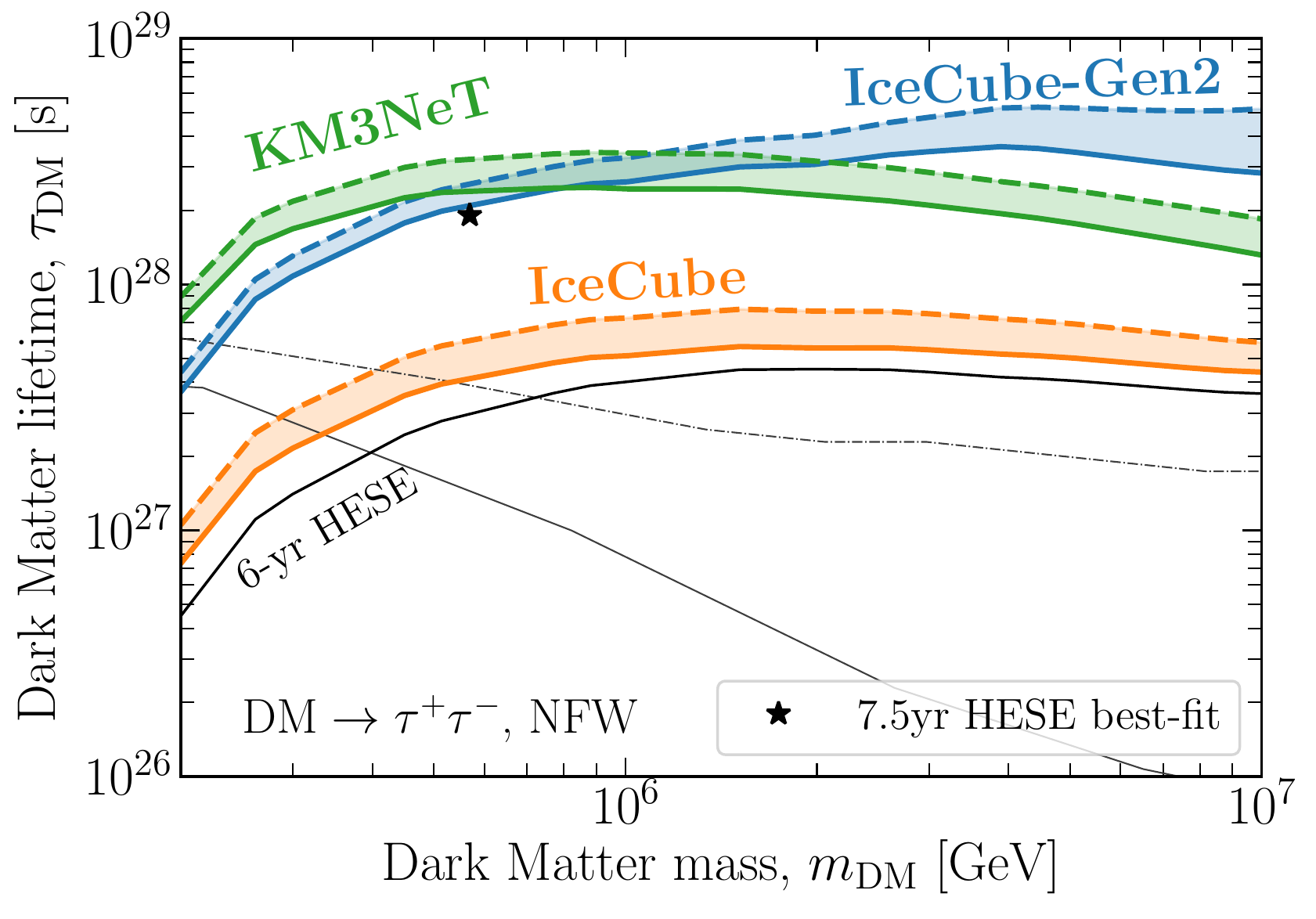}\hskip5.mm
\includegraphics[width=.48\textwidth]{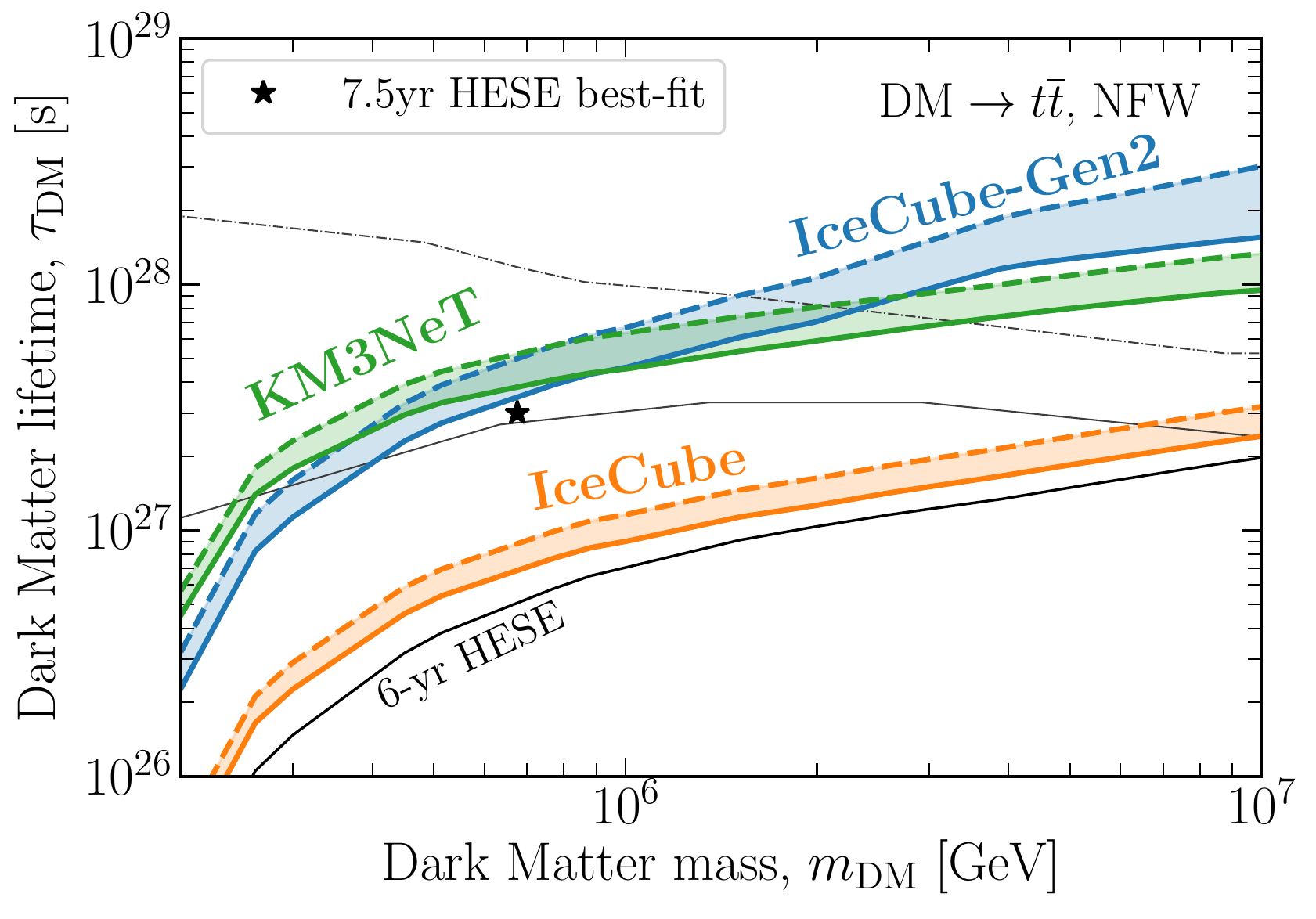}
\caption{{\bf Future sensitivity to decaying dark matter models.} Sensitivity at 95\% CL to DM lifetime as a function of DM mass after 10-year exposures of IceCube (orange), KM3NeT (green) and IceCube-Gen2 (blue) experiments. The bands represent the median (dashed lines) and 95\% (solid lines) conservative sensitivity from the Monte Carlo simulations. The current constraints from 6-year HESE data are shown with solid black lines. The left and right plots correspond to the $\tau^+\tau^-$ and $t\overline{t}$ channels, respectively. The black stars show the current best-fit of the DM component deduced by 7.5-year IceCube HESE data~\cite{Chianese:2019kyl}. The gamma-ray constraints are represented by light grey lines: solid from HAWC galactic halo searches~\cite{Abeysekara:2017jxs} and dot-dashed from Ref.~\cite{Cohen:2016uyg}.}
\label{fig:dec_sensitivity}
\end{figure}
\begin{figure}[h!]
\centering
\includegraphics[width=.48\textwidth]{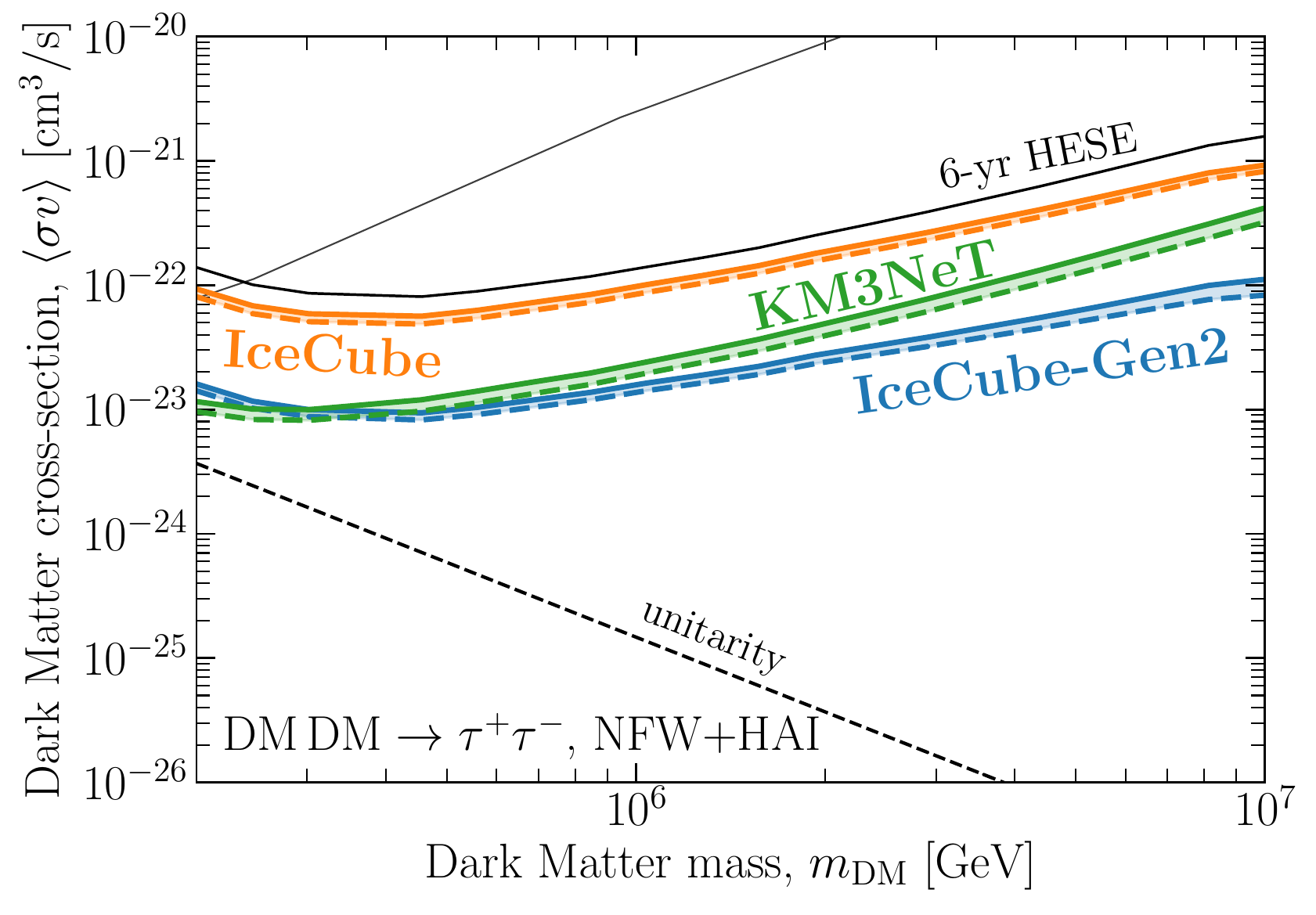}\hskip5.mm
\includegraphics[width=.48\textwidth]{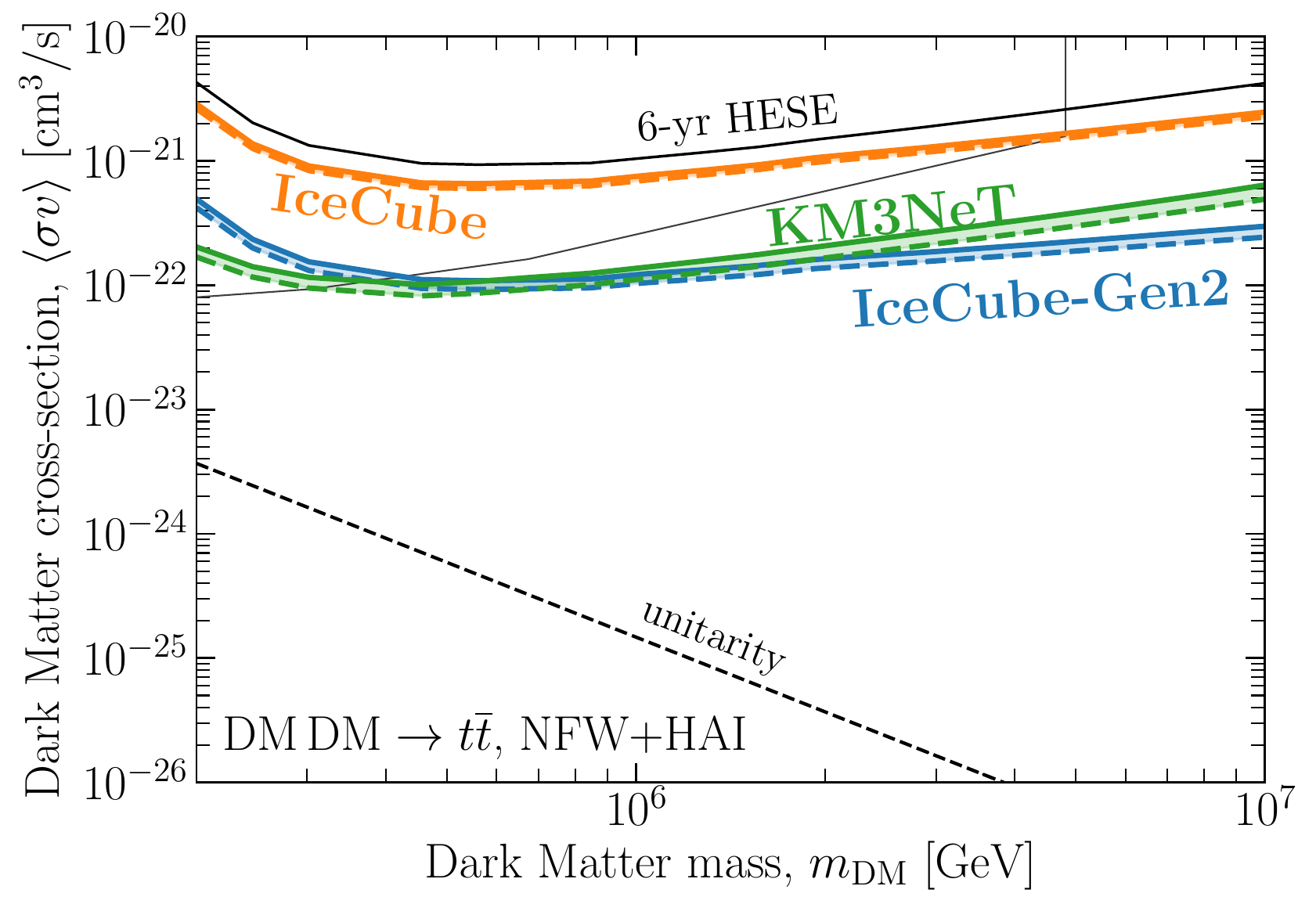}
\caption{{\bf Future sensitivity to annihilating dark matter models.} The plots are explained in the caption of Fig.~\ref{fig:dec_sensitivity}. In addition, the dashed black lines mark the unitarity limit on DM cross-section.}
\label{fig:ann_sensitivity}
\end{figure}

The main results of the present analysis are reported in Figs.~\ref{fig:dec_sensitivity} and~\ref{fig:ann_sensitivity}, where we show the future sensitivity of neutrino telescopes to decaying and annihilating DM models, respectively. We consider the channels into tau leptons ($\tau^+\tau^-$) and top quarks ($t\overline{t}$) as representative of different energy neutrino spectrum of the DM signals. The current constraints deduced from 6-year IceCube data are shown by the solid black lines. Other bounds shown are: gamma-ray searches from the galactic halo by HAWC~\cite{Abeysekara:2017jxs} (grey solid lines), global gamma-rays constraints from Ref.~\cite{Cohen:2016uyg} (grey dot-dashed lines), and the unitarity constraints on cross-section~\cite{Griest:1989wd} (black dashed lines). 

In the plots, the bands have been obtained by performing the APS analysis on simulated sky maps with different DM mass. They represent the sensitivity for an observed $p$--value of 0.05 from the median and upper 95\% values on the total DM events from the Monte Carlo simulations. Both KM3NeT and IceCube-Gen2 will probe a much bigger parameter space of DM models. In particular, KM3NeT turns out to be more sensitive to DM models with a low DM mass, as also shown in Fig.~\ref{fig:limits}. However, for DM masses larger than PeV, the sensitivity of KM3NeT is weaker than the one of IceCube-Gen2. This is due to two effects. At high energies, there is an enhancement in the expected number of shower events in IceCube thanks to the Glashow resonance of anti-electron neutrinos. This is not the case for KM3NeT, as it is based only on track-like events related to CC interactions of muon neutrinos. Moreover, KM3NeT will observe the galactic center through the Earth. While at low energies this is important to reduce the atmospheric (muon) background, at high energies the neutrino flux is reduced due to Earth absorption. On the other hand, the galactic center is observed by IceCube through down-going neutrinos that do not pass through the Earth.

Remarkably, KM3NeT and IceCube-Gen2 will probe a much bigger parameter space of DM models.In particular, they will be sensitive to the present 7.5-year HESE best-fit of the additional decaying DM component~\cite{Chianese:2019kyl}, shown with black stars in Fig.~\ref{fig:dec_sensitivity}. We emphasize that this result relies only on the angular information contained in the neutrino sky maps. This makes the APS analysis very robust against any potential degeneracies in the neutrino energy spectrum expected from astrophysical sources and DM particles. It is worth observing that for leptophilic channels the current best-fit of the DM component is not yet excluded by gamma-ray constraints.

\section{Discussion \label{sec:discussion}}

The results in the previous section have been obtained for benchmark models. However, other DM models can lead to different angular patterns and therefore lead to different exclusion limits. We thus discuss in detail how the sensitivity depends on DM properties like the channel, the DM mass, the galactic halo profile, and the boost factor using 10 years of IceCube-Gen2 exposure.

\begin{figure}[t!]
\centering
\includegraphics[width=.32\textwidth]{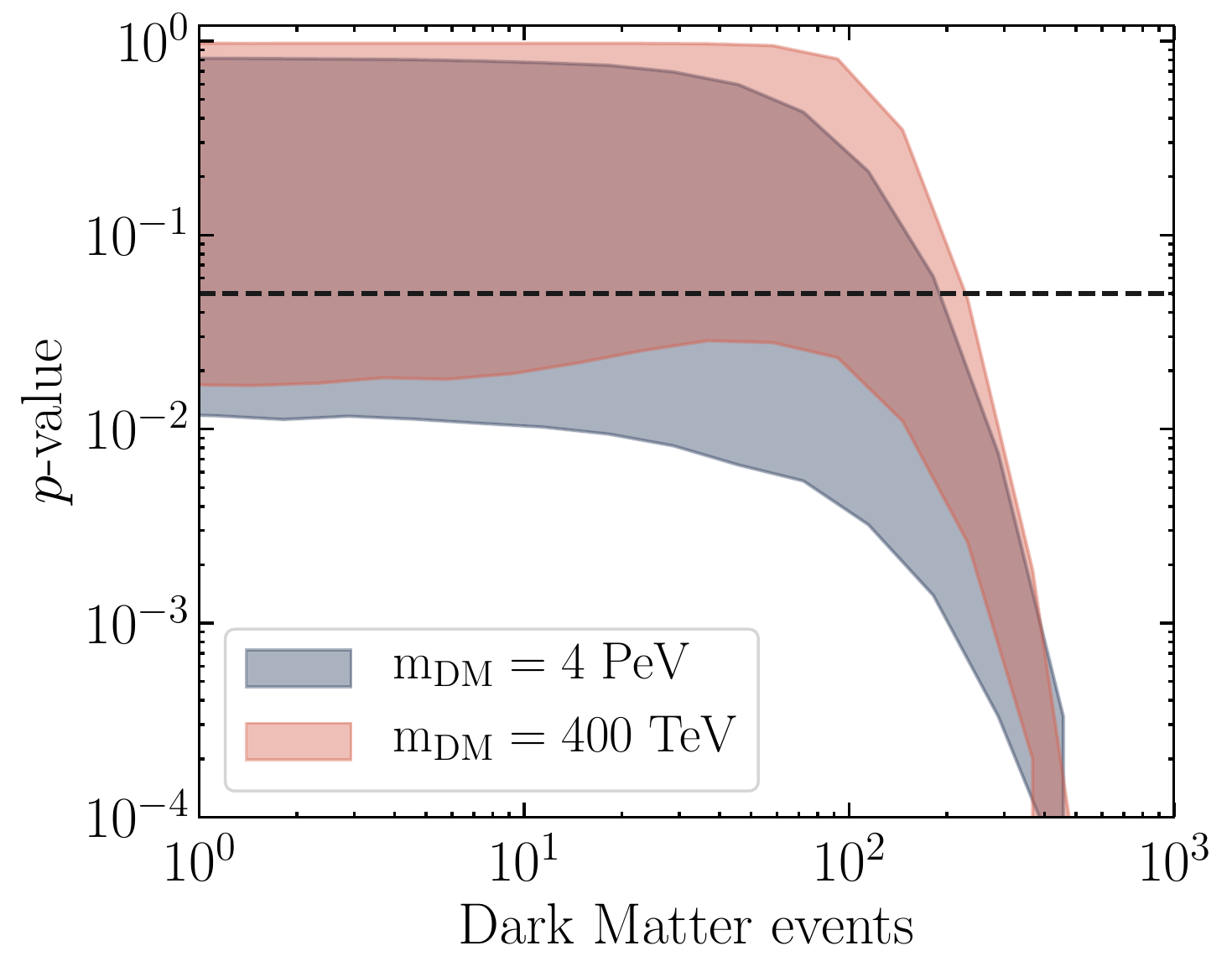}\hfill
\includegraphics[width=.32\textwidth]{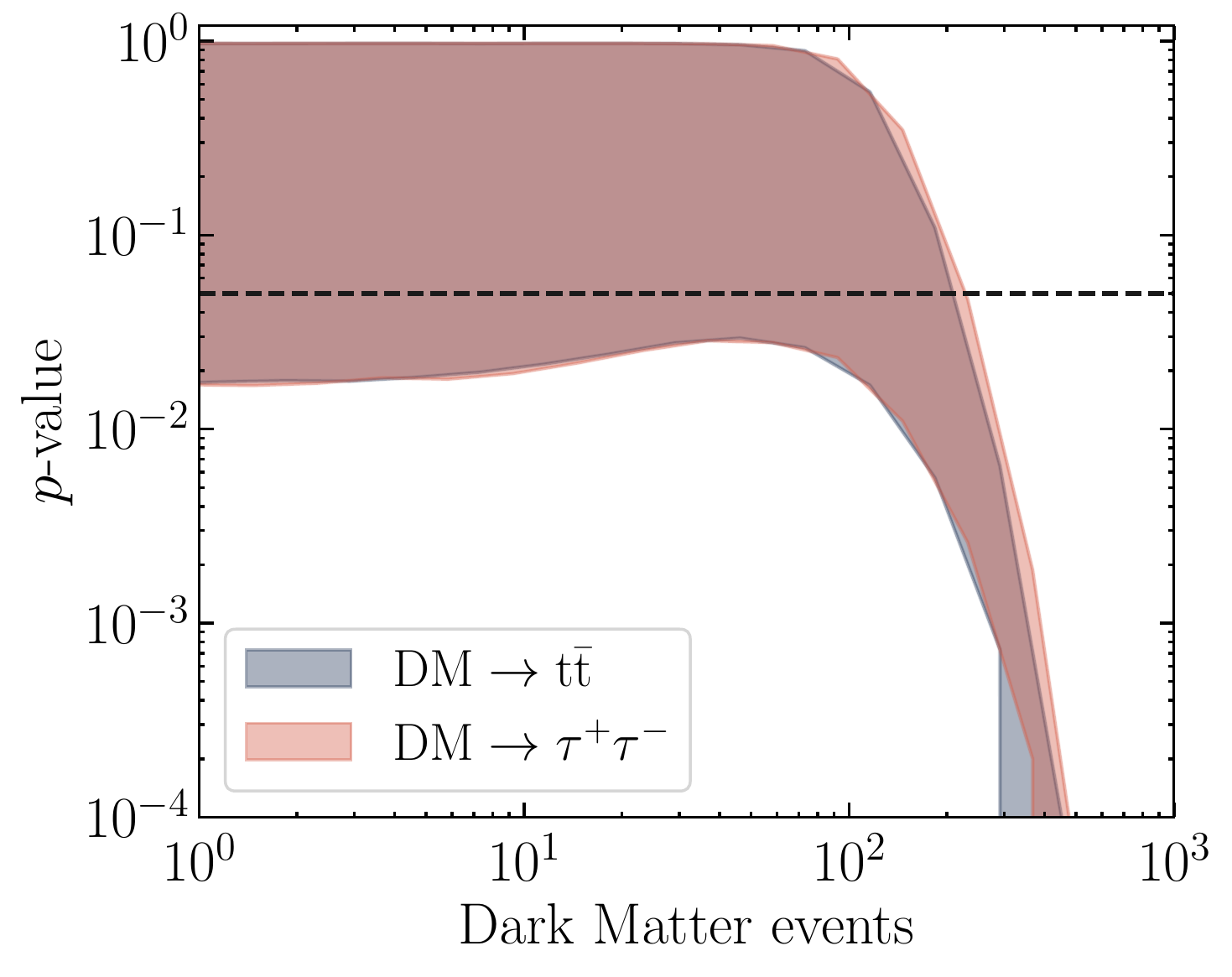}\hfill
\includegraphics[width=.32\textwidth]{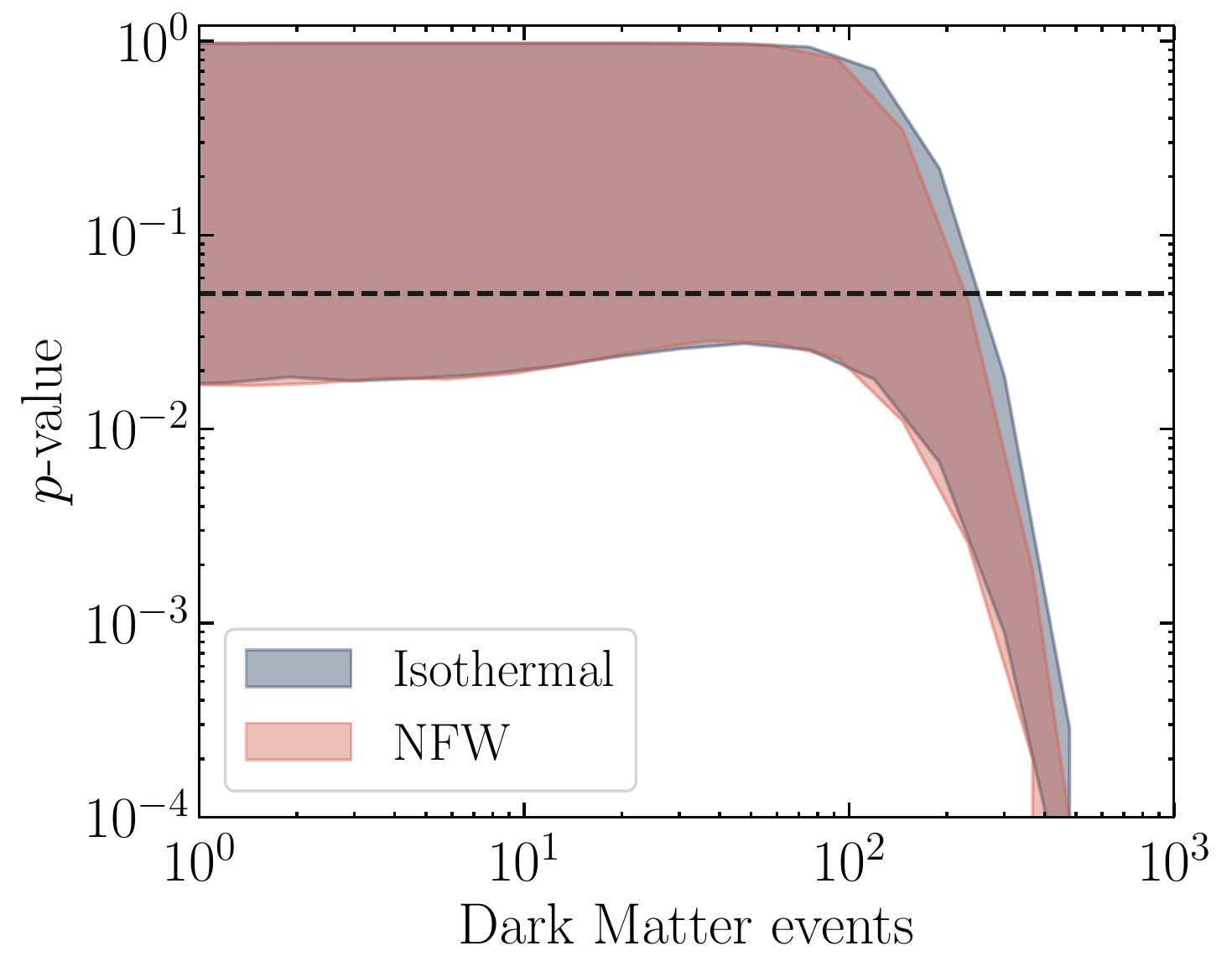}
\caption{\textbf{Sensitivity test for different decaying dark matter models.} Each panel shows the $p$--value band covering 95\% Monte Carlo simulations for the benchmark model ($m_{\rm{DM}}=400~\rm{TeV}$, ${\rm DM} \rightarrow\tau^- \tau^+$, NFW profile) in red together with changes to: $m_{\rm DM} = 4~$PeV (left), the channel ${\rm DM} \rightarrow t \overline{t}$ (middle) and the Isothermal density profile (right). The $p$-value bands correspond to 10 years of IceCube-Gen2 exposure.}
\label{fig:decay}
\end{figure}

For the case of decaying DM scenario, as shown in Fig.~\ref{fig:pval}, we have previously assumed $m_{\rm{DM}}=400~$TeV, $\rm DM \rightarrow\tau^- \tau^+$, and the NFW profile for the galactic halo. Figure~\ref{fig:decay} shows the 95\% contour $p$--value bands, where we compare in each panel the results for the benchmark parameters together with a parameter that we want to test. The left panel shows two different DM masses with $m_{\rm{DM}}=400~\rm TeV$ and $m_{\rm{DM}}=4~\rm PeV$, the middle panel the two channels of ${\rm DM} \rightarrow \tau^- \tau^+$ and ${\rm DM} \rightarrow t \overline{t}$, and the right panel the isothermal density profile together with the NFW profile. We find that the exclusion limits on the total DM events do not significantly change in all the cases. However, when plotted against the lifetime, the constraints change. For instance, the $\tau$--channel produces more events than the $t$--channel, and has therefore stronger constraints on the lifetime as shown in Fig.~\ref{fig:dec_sensitivity}. This allows one to set almost model-independent limits on the total DM events and then translate them into constraints on DM lifetime for different DM models. Concerning the DM halo (right panel), the agreement between NFW and Isothermal profiles is a result of similar ratio between galactic and extra-galactic DM events, whithin the angular resolution considered in this analysis.
\begin{figure}[t!]
\centering
\includegraphics[width=.32\textwidth]{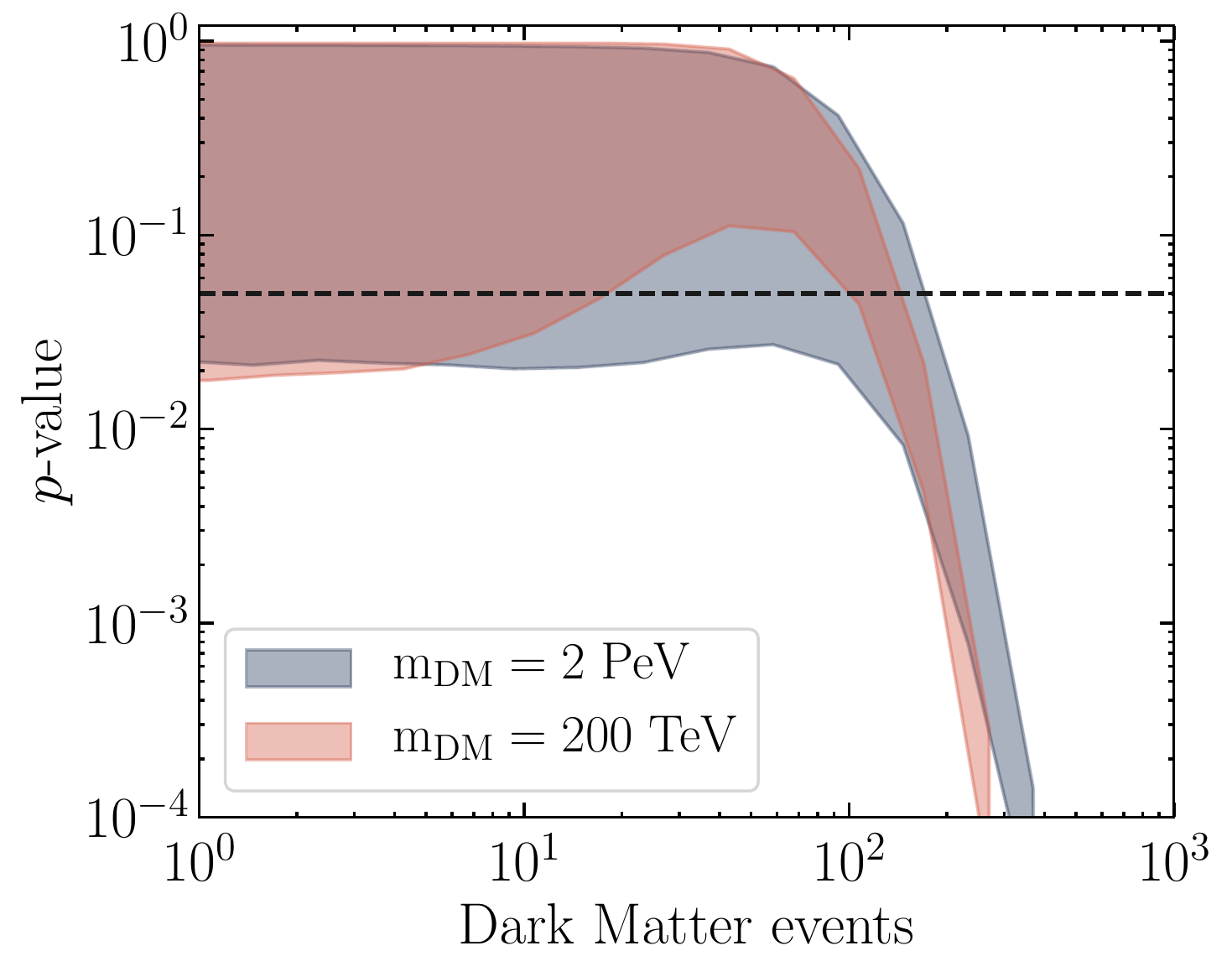}\hfill
\includegraphics[width=.32\textwidth]{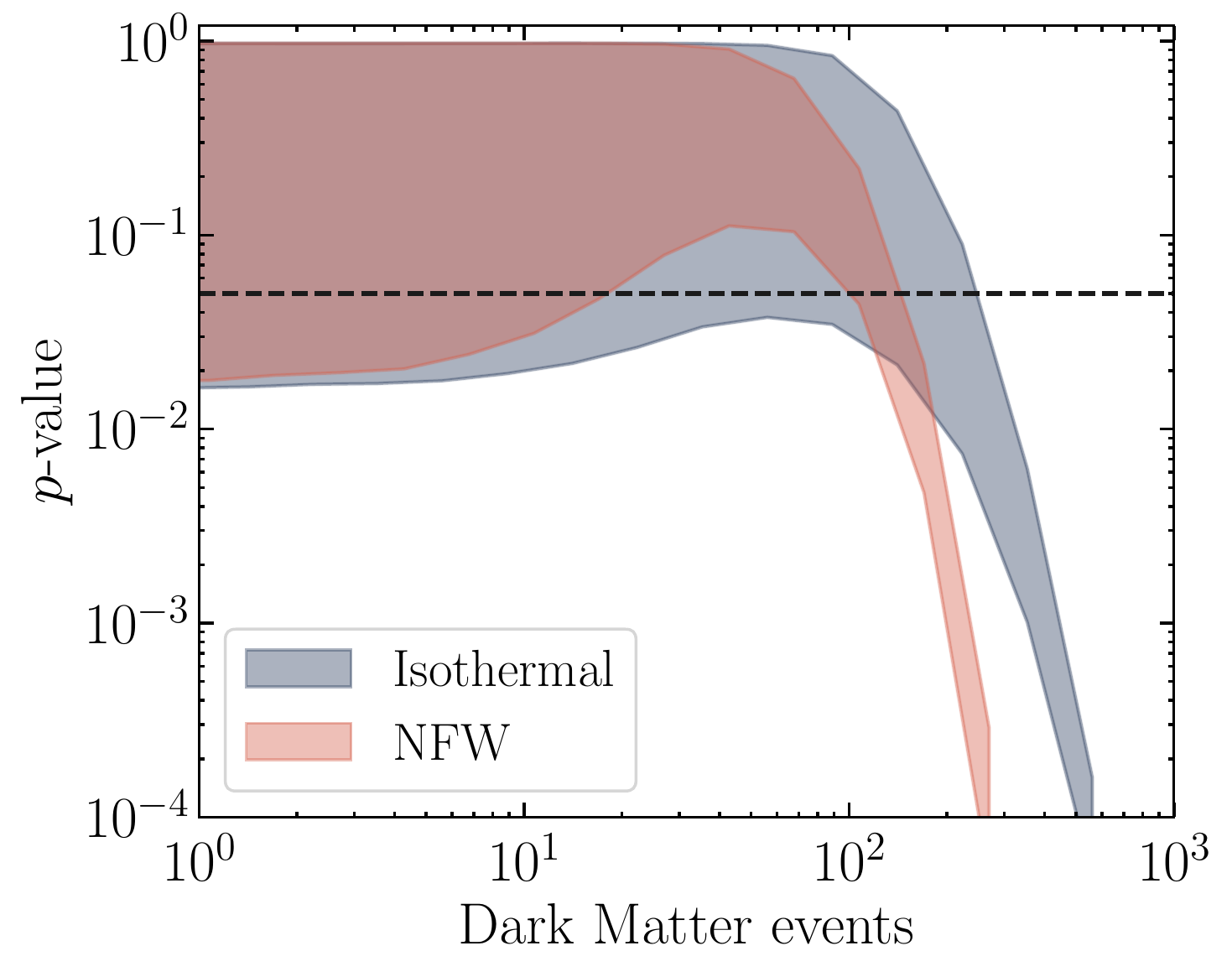}\hfill
\includegraphics[width=.32\textwidth]{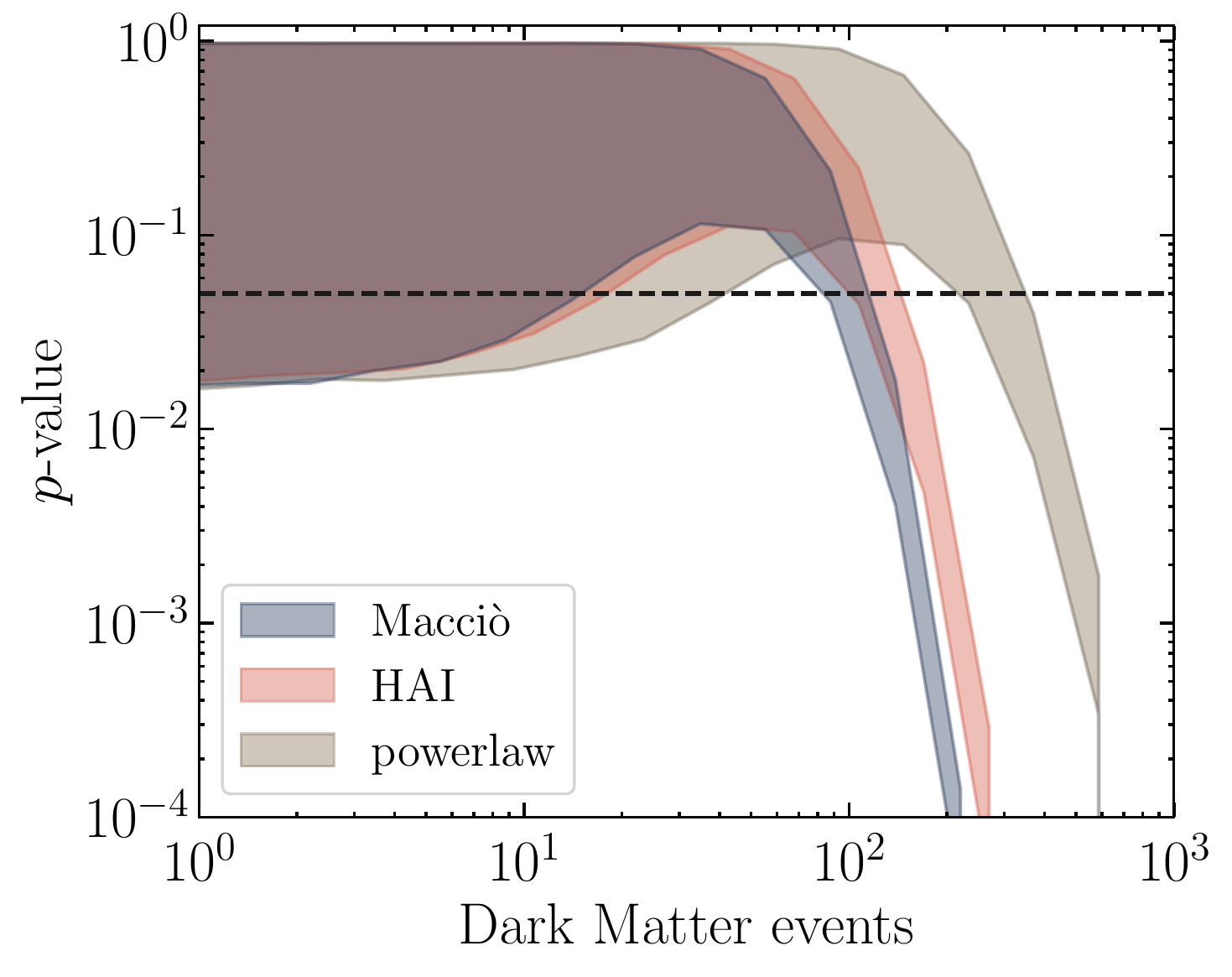}
\caption{\textbf{Sensitivity test for different annihilating dark matter models.} Each panel shows the $p$--value band covering 95\% Monte Carlo simulations for benchmark model ($m_{\rm{DM}}=200~\rm{TeV}$, ${\rm DM}\,{\rm DM} \rightarrow \tau^- \tau^+$, NFW profile, HAI boost factor) in red together with changes to: $m_{\rm DM} = 2~$PeV (left), the Isothermal density profile (middle), and the power-law and Macciò boost factors (right). The $p$-value bands correspond to 10 years of IceCube-Gen2 exposure.}
\label{fig:Annihilating}
\end{figure}

We perform the same tests for annihilating DM scenario for which we previously assumed the following benchmark model: $m_{\rm{DM}}=200~\rm TeV$, ${\rm DM} \, {\rm DM} \rightarrow \tau^- \tau^+$, NFW halo profile, and the HAI boost factor~\cite{Hiroshima:2018kfv}. Figure~\ref{fig:Annihilating} shows the 95\% contour bands of the $p$--value results, where we vary the DM parameters. As before, the sensitivity on the total DM events does not significantly depends on the channel, while considering larger DM masses reduces the limits by roughly 18\% (see for example the left panel). On the other hand, considering an isothermal density profile instead of the NFW one makes the constraints 70\% weaker, as shown in the middle panel.  This is due to the NFW density profile being denser in the inner center than the isothermal profile (see Fig.~\ref{fig:DM_properties}), and to the fact that in the case of annihilation the number of neutrino events scales stronger with the density than for decaying DM. Finally, the right panel shows the impact on the constraints for three different boost factors, namely the HAI, power-law and Macciò models (see Fig.~\ref{fig:DM_properties}). The boost factor is very important in setting constraints from angular information, since the ratio between galactic and extragalactic DM events strongly depends on it. In particular, the constrains with the power-law boost factor are weaker by a factor of 2.4 (140\%), while the ones for the Macciò model are stronger by a factor of 0.8 (20\%). It is worth observing that this difference has a direct impact on the sensitivity reported in Fig.~\ref{fig:ann_sensitivity}, which can be simply rescaled for different DM models.
\begin{figure}[t!]
\centering
\includegraphics[width=.4\textwidth]{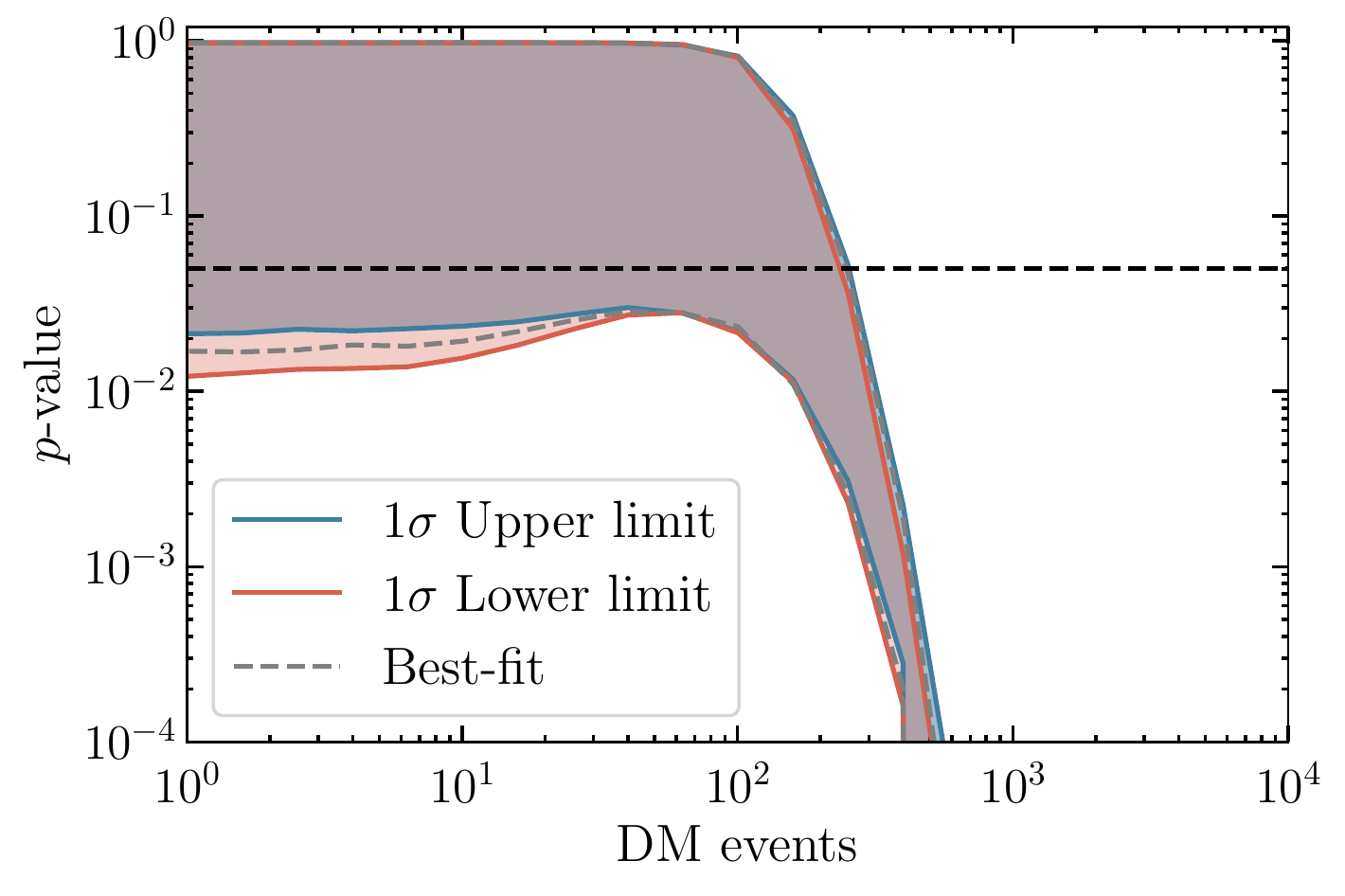}
\caption{\textbf{Uncertainty test for the astrophysical component} We test the robustness of our analysis against the $1\sigma$ upper and lower uncertainty of the astrophysical power-law obtained from the TG data sample. Shown are the 95\% contour bands of $p$--values from Monte Carlo simulations using benchmark model of decaying DM.}
\label{fig:Astro}
\end{figure}

Finally, in Fig.~\ref{fig:Astro} we discuss the impact of the statistical uncertainty affecting the best-fit for the astrophysical power-law from the TG data sample. In particular, we test if our results are robust for the $1\sigma$ statistical uncertainty on the normalization and spectral index of the observed spectrum and show here the $p$--value result for the benchmark model for decaying DM particles. The best-fit is shown as the grey shaded area delimited by dashed lines, and in blue and orange are the $1\sigma$ upper and lower limit. As can be clearly seen, in all the cases the corresponding bounds on the total DM events are the same. 

\section{Conclusions \label{sec:conclusions}}

In this paper, we have investigated the two-component interpretation of the high-energy neutrino flux observed by IceCube. The possibility of two different components contributing to the observed neutrino flux in addition to the atmospheric background has been proposed to solve the slight tension between IceCube HESE and TG data. In particular, we have focused on the scenario where one of the two neutrino components is originated by decaying or annihilating dark matter. In order to constrain the total number of neutrino potentially related to dark matter, we have studied the angular power spectrum of neutrino sky maps for different scenarios: the null hypothesis of atmospheric neutrino background and an astrophysical power-law component, and the signal hypothesis that includes a dark matter component. While the former is nearly isotropic above 60~TeV, the latter is expected to correlate with the galactic halo of the Milky Way. By means of Monte Carlo simulations and a $\chi^2$ analysis, we have provided the current constraints on dark matter properties (lifetime and cross-section as a function of DM mass) deduced from 6-year IceCube HESE shower data. Moreover, we have reported the future sensitivity to a dark matter signal after 10 years of observations with IceCube and with next-generation neutrino telescopes as KM3NeT and IceCube-Gen2. KM3NeT has been found to be more sensitive to a dark matter component for low dark matter masses, thanks to the better sensitivity to the galactic center at energies below the threshold of Earth absorption. Finally, this analysis has shown that both KM3NeT and IceCube-Gen2 will be able to firmly probe the current 7.5-year HESE best-fit of the dark matter component by exploiting angular information only. This result is of paramount importance since it highlights a feasible and solid way to distinguish a dark matter signal to neutrino fluxes produced by potentially hidden astrophysical sources.

\section*{Acknowledgments}
We thank Rasa Muller for useful discussions. This work was supported by GRAPPA Institute, University of Amsterdam. SA also acknowledges partial support by JSPS KAKENHI Grant Numbers JP17H04836, JP18H04578, and JP18H04340.

\bibliography{references}

\end{document}